\documentclass[twocolumn,english,prx,showpacs]{revtex4-1}
\usepackage[colorlinks=true,urlcolor=blue,citecolor=blue,linkcolor=blue]{hyperref}
\usepackage{natbib}
\usepackage[T1]{fontenc}
\usepackage[latin9]{inputenc}
\usepackage{amssymb}
\usepackage{graphicx}
\usepackage{amsmath,color}
\usepackage{mathrsfs}
\usepackage{float}
\usepackage{indentfirst}
\usepackage{babel}
\usepackage{color}
\makeatletter

\begin{document}

\title{Type II Weyl Semimetals}

\author{Alexey A. Soluyanov$^{1}$}
\author{Dominik Gresch$^{1}$}
\author{Zhijun Wang$^{3}$}
\author{QuanSheng Wu$^{1}$}
\author{Matthias Troyer$^{1}$}
\author{Xi Dai$^{2}$}
\author{B. Andrei Bernevig$^{3}$}
\affiliation{${^1}$Institute for Theoretical Physics, ETH Z\"urich, 8093 Z\"urich, Switzerland}
\affiliation{${^2}$Institute of Physics, Chinese Academy of Sciences, Beijin 100190, China}
\affiliation{${^3}$Department of Physics, Princeton University, New Jersey 08544, USA}

\date{\today}

\maketitle

{\bf  Fermions in nature come in several types: Dirac, Majorana and Weyl are theoretically thought to form a complete list. Even though Majorana and Weyl fermions have for decades remained experimentally elusive, condensed matter has recently emerged as fertile ground for their discovery as low energy excitations of realistic materials ~\cite{Wan-PRB11, Volovik-book, Weng-PRX15, Huang-NatComm15, Xu-Science15, Lv-PRX15}. Here we show the existence of yet another particle - a new type of Weyl fermion - that emerges  at the boundary between electron and hole pockets in a new type of Weyl semimetal phase of matter. This fermion was missed by Weyl in 1929 due to its breaking of  the stringent  Lorentz symmetry of high-energy physics. Lorentz invariance however is not present in condensed matter physics, and we predict that an established material, WTe$_2$, is an example of this novel type of topological semimetal hosting the new particle as a low energy excitation around a type-2 Weyl node. This node, although still a protected crossing, has an open, finite-density of states Fermi surface, likely resulting in a plethora physical properties very different from those of standard point-like Fermi surface Weyl points.}

\begin{figure}
\begin{center}
\includegraphics[width=\columnwidth]{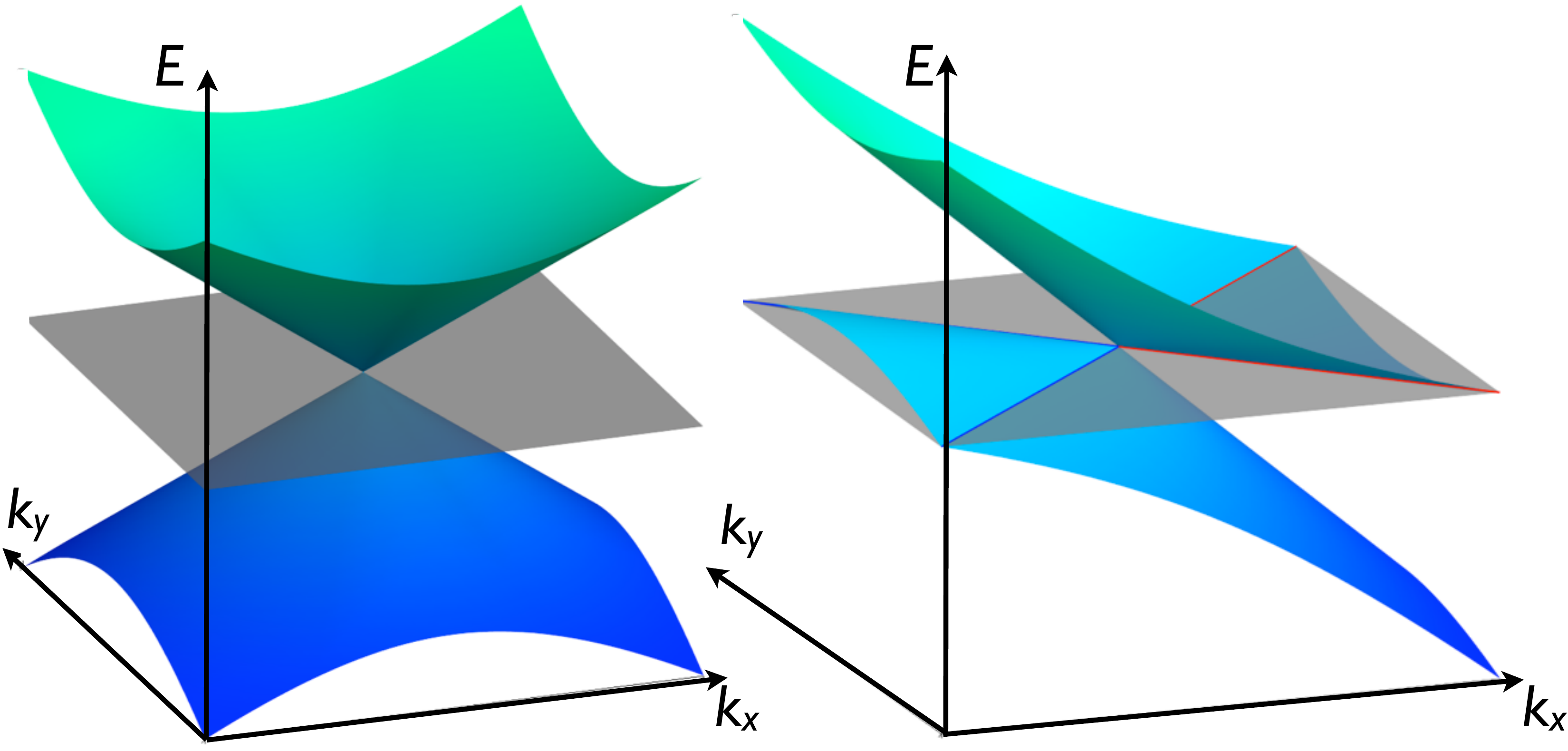}
\end{center}
\caption{Possible types of Weyl semimetals. Left panel: Type I Weyl point with point-like Fermi surface. Right panel: Type II Weyl point is the touching point between electron and hole pockets.}
\label{fig:intro}
\end{figure}

 Metals  can exhibit non-trivial topological features~\cite{Volovik-book}. Of these, the ones with vanishingly small density of states at the Fermi level, called semimetals (SM), stand out. For these, a distinction between topologically protected surface and bulk metallic states can still be made, and their Fermi surfaces (FS) exhibit a topological (T) characterization. Two kinds of TSM have attracted special attention: Dirac and Weyl SMs. In these materials a linear crossing of two (Weyl) or four (Dirac) bands occurs at the Fermi level (see left panel of Fig.~\ref{fig:intro}). The effective Hamiltonian for these crossings is given by the Weyl or gapless Dirac equation respectively. The Weyl crossings are protected from gapping, a statement of the massless nature of the Weyl fermion. In the following we limit the discussion to Weyl crossings only, although our results also hold for  Dirac crossings.

The appearance of Weyl points (WPs) is only possible if the product of parity and time-reversal (TR) is not a symmetry of the structure. When present, a WP acts as a topological charge - either a source or a sink of Berry curvature. A FS enclosing a WP has a well-defined Chern number, corresponding to this point's topological charge. Since the net charge has to vanish in the entire Brillouin zone (BZ), WPs always come in pairs; WPs are stable to weak perturbations and can only be annihilated in pairs of opposite charge. A large number of unusual physical phenomena are associated with Weyl TSM, including the existence of open Fermi arcs on the surface FS~\cite{Wan-PRB11, Silaev-PRB12}, and different  magnetotransport anomalies~\cite{Nielsen-PLB83, Zyuzin-PRB12, Hosur-crp13, Volovik-JETPL14, Zhang-arx15, Xiong-arx15, Huang-arx15b}.  

Weyl SMs with broken TR symmetry have been predicted to exist in several materials~\cite{Wan-PRB11, Xu-PRL11, Burkov-PRL11}, but experimental verification has been lacking. More recently, the Weyl SM was predicted to exist in inversion-breaking single crystal nonmagnetic materials of the TaAs class~\cite{Weng-PRX15, Huang-NatComm15}; this prediction was soon verified experimentally~\cite{Xu-Science15, Lv-PRX15}. 

To present, Weyl SMs have been thought to have a point-like FS at the WP. We refer to these as type-I WPs (WP1), to contrast them with the new type-II WPs (WP2) 
that exist at the boundary of electron and hole pockets, as illustrated in the right panel of Fig.~\ref{fig:intro}. We discuss general conditions for WP2s to appear, and present evidence that WTe$_2$, a material recently reported~\cite{Ali-Nature14} to have the largest known to date never-saturating magnetoresistance, is an example of the the new type of TSM hosting eight WP2s. They come in two quartets located $0.052$~eV and $0.058$~eV above the Fermi level. We present topological arguments proving the existence of the novel TSM phase in this material. Doping-driven topological Lifshits transitions, characteristic of WP2s, as well as emerging Fermi arcs at the surface, are presented.

We start by considering the most general Hamiltonian describing a WP
\begin{equation}
H(\bf k)=\sum_{ij} k_i A_{ij}\sigma_j
\end{equation}
where $A_{ij}$ is a 3$\times$4 matrix, with $i=x,y,z$, and $j=0,x,y,z$ being indices of the 2$\times$2 unit and three Pauli matrices. The spectrum is given by 
\begin{equation}
\varepsilon_{\pm}({\bf k})=\sum_{i=1}^3 k_iA_{i0}\pm \sqrt{\sum_{j=1}^3 (\sum_{i=1}^3 k_iA_{ij})^2}=T({\bf k})\pm U({\bf k})
\end{equation}
The kinetic part of energy, $T({\bf k})$, also linear in momentum, tilts the cone-like spectrum. Breaking the Lorentz invariance of the Dirac Weyl fermions in quantum field theory, this tilt was previously considered unimportant. Since Lorentz invariance does not need to be respected in condensed matter, its inclusion is important and leads to a finer classification of distinct FSs in a one-to-one correspondence with the theory of quadric surfaces. The conclusion: there are exactly two distinct types of WPs~\cite{genroute} (see Supplementary Material). 

For a direction in reciprocal space, where $T$ is dominant over $U$, the tilt becomes large enough to produce touching electron and hole pockets, rather than a point-like FS. Thus, the condition for a WP to be of type II is that there exists a direction $\hat{\bf k}$, for which $T(\hat{\bf k})>U(\hat{\bf k})$. If such a direction does not exist, the WP is of Type I. The clear \emph{qualitative} distinction in FSs of the two types of WPs leads to drastic differences in the thermodynamics and response to magnetic fields~\cite{LLs}. In particular, in sharp contrast to WP1, which exhibits a chiral anomaly~\cite{Nielsen-PLB83} for any direction of the magnetic field, the chiral anomaly appears in a WP2 when the direction of the magnetic field is  within a  cone where $T({\bf k})$ dominates. If the field direction is outside of this cone the Landau level spectrum is gapped and has no chiral zero mode~\cite{LLs} (see Supplementary Material).

On the lattice, the no-go theorem~\cite{Nielsen-NPB81} guarantees that Weyl fermions appear in pairs with opposite Chern numbers. Since the Chern number of a WP is not changed by the kinetic term, WPs of different type can be chiral/anti-chiral partners of each other. The number of WPs of certain type could be odd, but the total number of WPs  has to be even (e.g. there can be one WP1 and one WP2).     

We now proceed to describe WTe$_2$, a material we identified to host the new WPs. The crystal structure of WTe$_2$ is orthorhombic with space group $Pnm2_1$ ($C_{2v}^7$). Its primitive unit cell contains four formula units. The atomic structure is layered, with single layers of W separated from each other by Te bilayers and stacked along the $z$-axis (see Supplementary Material). The distance between adjacent W atoms is considerably smaller along the $x$-axis, creating strong anisotropy. The unit cell has two reflection symmetries: mirror in the $yz$-plane, and a glide plane formed by a reflection in the $xz$-plane followed by a translation by $(0.5, 0, 0.5)$. Combined they form a non-symmorphic two-fold rotation $C_2$, which is important in the following symmetry arguments.
 
\begin{figure}
\begin{center}
\includegraphics[width=\columnwidth]{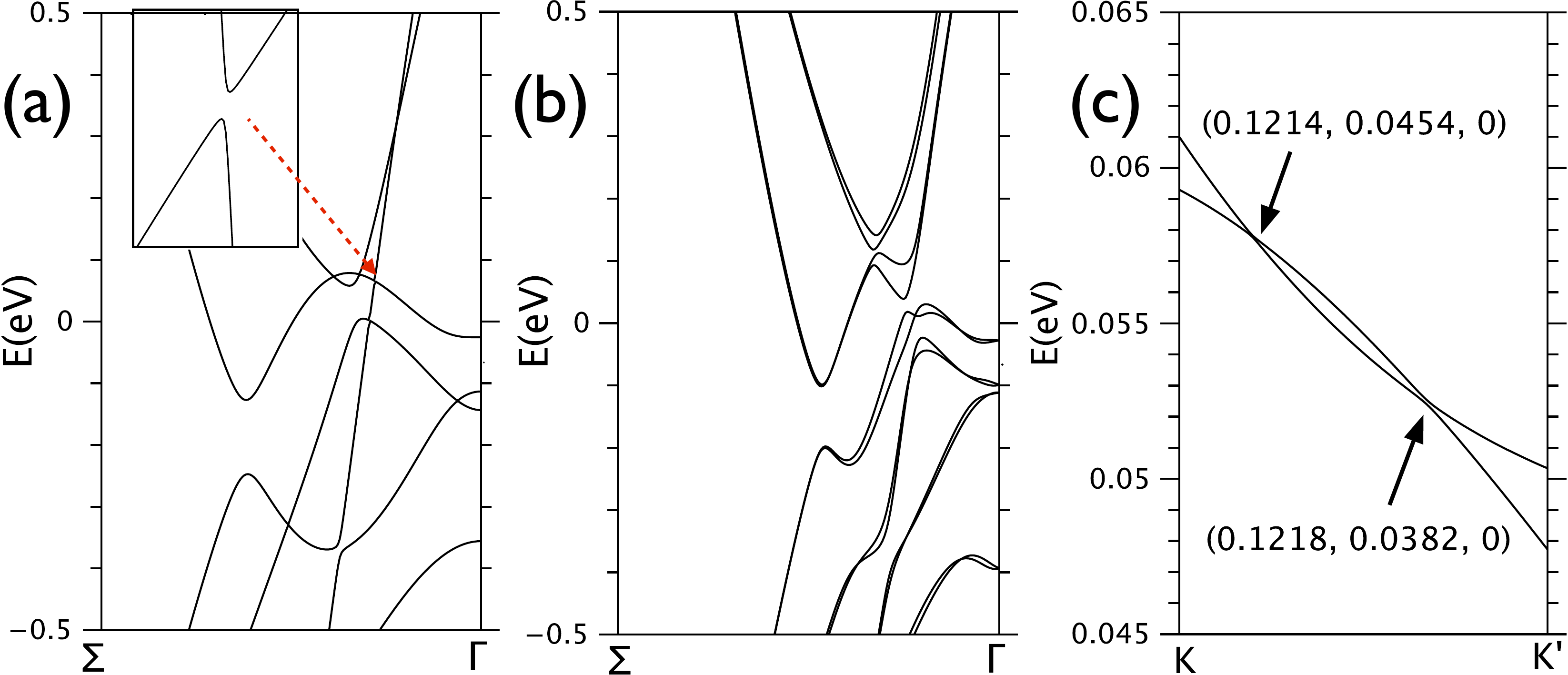}
\end{center}
\caption{Band structure of WTe$_2$. Panel (a): Band structure of WTe$_2$ without SOC. A fraction of $\Gamma$-$X$ segment is shown: the point $\Sigma$ has coordinates $(0.375, 0, 0)$. Band gap of $\approx 1$meV is shown in the inset signaling a gapless point nearby. Panel (b): Band structure of WTe$_2$ with SOC. Panel (c): One of the four pairs of WPs. Their locations are designated in reduced coordinates. Zero energy is set to $E_{\rm F}$ in all panels.}
\label{fig:bstrs}
\end{figure}

The result of band structure calculations (see Supplementary Material) without spin-orbit coupling (SOC) is shown in Fig.~\ref{fig:bstrs}(a) along the $\Gamma$-$X$ direction, where an intermediate point $\Sigma=(0.375, 0, 0)$ is introduced. Apart from electron-hole pockets, 16 WPs per spin are found in WTe$_2$ in the absence of SOC  (not shown in Fig.~\ref{fig:bstrs}(a)). Half of these points occur at points of low symmetry at $k_z\neq 0$. The other half of SOC-free Weyls appear in the $k_z=0$ plane, where the product of TR and $C_2$ ($C_{2T}=C_2*T$) forms a little group. Generically degeneracies on high symmetry planes are forbidden; however, due to the $C_{2T}$ symmetry two-fold degeneracies are locally stable at points in the $k_z=0$ plane~\cite{genroute}. On the $\Gamma$-$X$ line the spectrum is generally gapped with a band gap of $\approx 1$meV, separating valence and conduction bands; see Fig.~\ref{fig:bstrs}(a).

Accounting for spin, but without SOC, bands become doubly degenerate corresponding to opposite spin projections. This doubles the topological charge of each Weyl, since by $\mathrm{SU}(2)$ symmetry WPs corresponding to opposite spins have identical topological charge. Infinitesimal SOC cannot gap them, giving a general recipe to search for Weyl semimetals~\cite{genroute}.

In WTe$_2$, however, SOC is not small. When turned on, it preserves electron-hole pockets, but significantly changes the structure of WPs. At intermediate SOC WPs can move, emerge or annihilate in pairs of opposite chirality. At full SOC all WPs at $k_z\neq 0$ are annihilated. In the $k_z=0$ plane, double degeneracies at isolated $k$-points are still allowed by symmetry~\cite{genroute}. Eight such gapless points are found, formed by the topmost valence and lowest conduction bands at full SOC. A pair of such points is shown in Fig.~\ref{fig:bstrs}(c). The other three pairs are related to this one by reflections. Energetically both points are located only slightly ($0.052$ and $0.058$~eV) above the Fermi energy. See Supplementary Material for details.

Establishing degeneracies of bands (WP) through ab-intio is prone to finite size effects: one can imagine an unfortunate scenario where a degeneracy point actually turns out to have a minuscule gap upon further increase in computing resolution. To rigorously establish the degeneracies of WPs, we have performed a multitude of error-free tests that involve computing topological indices. These are presented in the Supplementary Material. We have computed the topological charge ($\pm 1$) of each Weyl point found through an extension of the Willson loop eigenvalue method ~\cite{Soluyanov-PRB11-b,Yu-PRB11} to type II Weyls. We have then computed the $\mathbb{Z}_2$ topological indices on a multitude of planes (including both standard and non-standard geometries) in the Brillouin zone. In total, these tests not only proved the existence of Weyl nodes but also were more than sufficient to elucidated the structure of the Berry flux connection between Weyls and of the Fermi arcs on the surface of $WTe_2$. The Fermi arc structure is consistent to the one presented through ab-initio calculations below.

To check the new nature of the WPs, we obtained their dispersion from first-principles calculations and fitted it to the theoretical one derived by symmetry analysis (Supplementary Material). Keeping terms linear in $k_i$, the momentum relative to the position of the WP, we get:
\begin{equation}
\varepsilon_{\bf k}=Ak_x+Bk_y\pm \sqrt{e^2k_z^2+(ak_x+ck_y)^2+(bk_x+dk_y)^2}
\end{equation}
The values of the parameters $A$, $B$, $a$, $b$, $c$, $d$ and $e$ are given in the Supplementary Material. The kinetic part of the energy dominates along the line connecting the nearest WP (see Fig.~\ref{fig:bstrs}(c) ). We thus conclude that WTe$_2$ is a type II Weyl SM. 

\begin{figure}
\begin{center}
\includegraphics[width=\columnwidth]{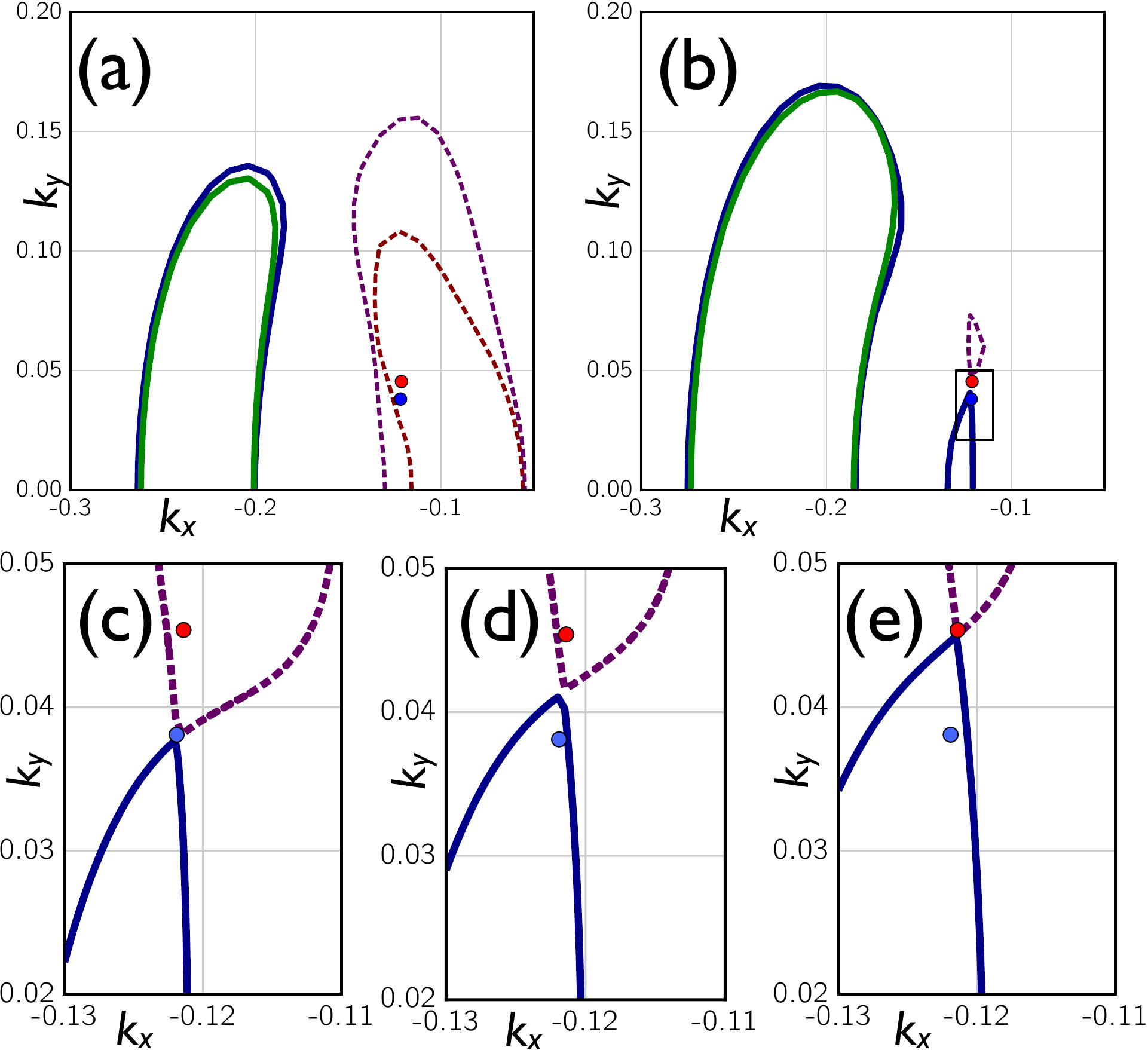}
\end{center}
\caption{Fermi surface at $k_z=0$. A part of the BZ is shown. (a). $E_{\rm F}=0$eV: electron (blue and green solid) and hole (red and magenta dashed) pockets come in pairs. (b). The structure of electron and hole pockets at higher energies ($E_{\rm F}=0.055$eV shown). There are four hole (1 shown) and 6 electron (halves of the 3 of them shown) pockets. The zoom of the framed region is shown in (c)-(e) for different values of $E_{\rm F}$. (c). $E_{\rm F}=0.052$eV is set to the lower energy Weyl point. A touching between $n$- and $p$- pockets occurs at the Weyl point. (d). $E_{\rm F}=0.055$eV is set to be in between the two Weyl points. The $n$- and $p$-pockets are disconnected. The hole pocket encloses a $C=+1$ Weyl point. The electron pocket encloses the $C=-1$ point and its mirror image (not shown); the net Chern number of this pocket is zero. (e). At the higher energy Weyl point $E_{\rm F}=0.058$eV electron and hole pockets touch again (shown). They reopen at larger $E_{\rm F}$ with zero Chern numbers.}
\label{fig:lifshits}
\end{figure}

We now discuss the FS topology and possible topological Lifshits transitions in this compound. The evolution of the FS obtained from first-principles calculations is shown in Fig.~\ref{fig:lifshits} for different values of $E_{\rm F}$. Due to reflection symmetries, only part of the $k_z=0$ cut of the FS is shown. For $E_{\rm F}=0$~eV the FS is formed of two pairs of electron and two pairs of hole pockets (8 pockets total), separated in momentum space. For each pair a larger surface completely encloses the smaller one, in agreement with experiments~\cite{Pletikosic-PRL14}. This is shown in Fig.~\ref{fig:lifshits}(a), where 4 halved (2 $n$ and 2 $p$) pockets are shown. The other halves are obtained by glide reflection $g_{xz}$;  remaining $k_x>0$ 4 pockets  are obtained by mirror  $m_{yz}$. All Fermi surfaces have zero Chern numbers in this case.

When $E_{\rm F}$ is raised, two additional electron pockets appear; the previously existing electron pockets persist. The hole pockets shrink quickly, two disappearing completely. Each of the other two split into two disconnected pockets. As a result, there are 6 electron- and 4 hole-pockets in total (see Fig.~\ref{fig:lifshits}(b) for illustration). For the Fermi level tuned to the first WP $E_{\rm F}=0.052$~eV, (corresponding to the addition of $\approx 0.064$ electrons per unit cell), each of the two newly appeared $n$-pockets touches two $p$-pockets at the positions of the WPs, as illustrated in Fig.~\ref{fig:lifshits}(c) for the part of $k_z=0$ plane. Further increase of $E_{\rm F}$ {\it disconnects} the electron and hole pockets again - see Fig.~\ref{fig:lifshits}(d) for $E_{\rm F}=0.055$~eV - but with changed topology: electron pockets still have zero Chern numbers, since they enclose two WPs of opposite charge, related by $g_{xz}$. The hole pockets have Chern numbers $\pm 1$. Topologies of the other $p$-pockets can be obtained by Chern-number flipping mirror and glide symmetries.  The pockets touch again (see Fig.\ref{fig:lifshits}(e)) when the Fermi level is tuned to the higher energy WP ($E_{\rm F}=0.058$~eV, with $\approx 0.079$ additional electrons per unit cell). Upon raising $E_{\rm F}$ further, the pockets disconnect again, and all FS Chern numbers become zero. 

 WTe$_2$ is close to a topological transition occurring when the closeby WP of opposite chirality annihilate each other. To facilitate the observation of topological Lifshits transitions, hydrostatic pressure can be applied. Neighboring WPs are pushed away from each other in $k$-space under compression.  In particular, a $0.5\%$ ($2\%$) compression increases the distance between the WPs from $~0.7\%$ to $~2\%$ ($~4\%$) of the reciprocal vector $|G_2|$ (see Supplementary Material for a thorough discussion of strain effects, including how to obtain only four Weyl points).  

\begin{figure}
\begin{center}
\includegraphics[width=\columnwidth]{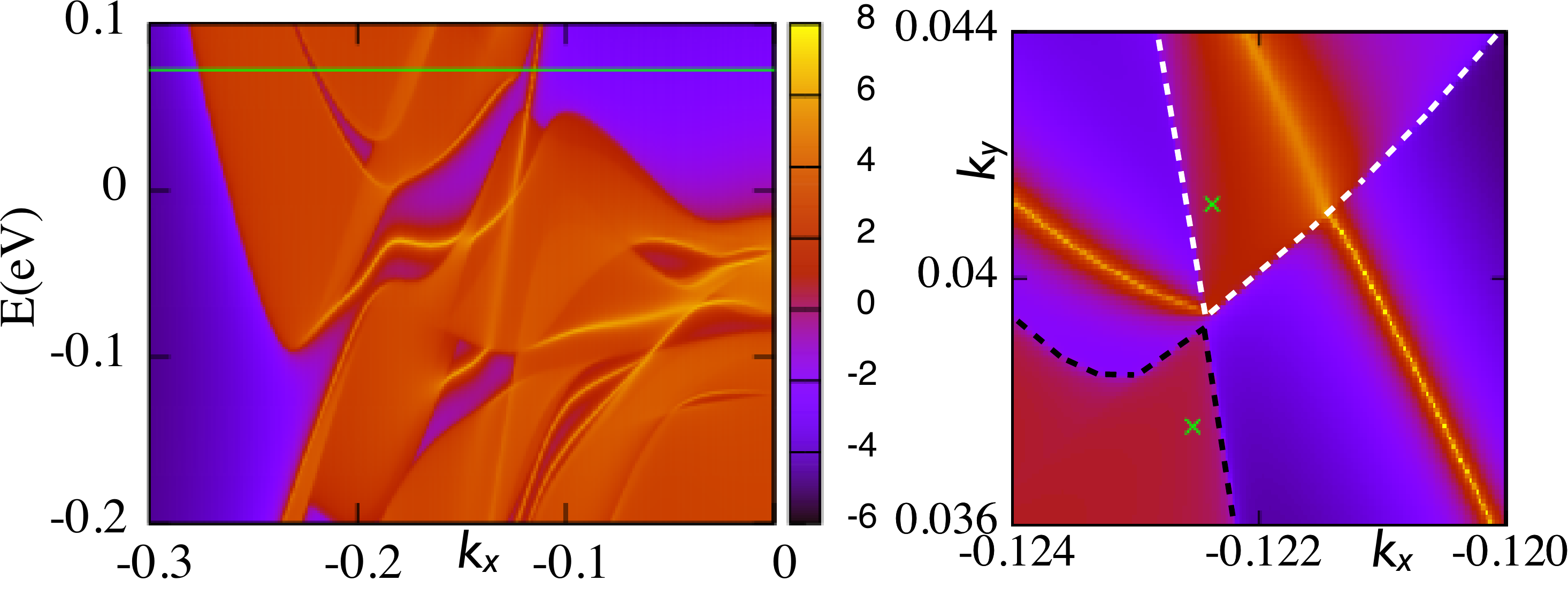}
\end{center}
\caption{Topological surface states. Left panel: spectral function of the $(001)$ surface. Fermi level is set in between the Weyl points (shown in green). Right panel: $(001)$ surface Fermi surface and a Fermi arc connecting $p$- and $n$-pockets. Green crosses mark the positions of Weyl points.}
\label{fig:surf}
\end{figure}

Finally, we discuss the topological surface states of WTe$_2$. Due to reflection symmetries, WPs of opposite chirality are projected on top of each other on the $(100)$ and $(010)$ surfaces which hence do not exhibit topologically protected surface states. For the $(001)$ surface all the WPs project onto distinct points, hence topological surface states appear. When $E_{\rm F}$ is tuned in between the WPs, the hole pocket has non-zero Chern number and a Fermi Arc emerges from it, connecting it to the WP of opposite Chern number inside the $n$-pocket. Fig.~\ref{fig:surf} illustrates the $(001)$ surface spectral function where surface states connecting electron and hole bands are clearly visible in the left panel. 
The fermi surface of these surface states is a topological Fermi arc (right panel of Fig.~\ref{fig:surf}) connecting the topological $p$-pocket (Fig.~\ref{fig:lifshits}(c-e)) to the $n$-pocket.  The other surface state crossing the $p$-pocket emerges from the $n$-pocket (not seen in the figure) and goes back into it, and thus can be pushed into the continuum of bulk states not to be seen at this $E_{\rm F}$ (see Supplementary Material).

Of other transition metal dichalcogenides another strong candidate material is MoTe$_2$~\cite{Brown-AC66} - reported to be a semimetal resembling pressurized WTe$_2$.  These can be used to explore new physical phenomena arising in this novel TSM phase.   

{\it Acknowledgements.} AAS, DG, QSW and MT acknowledge the support of Microsoft Research, the Swiss National Science Foundation through the National Competence Center in Research MARVEL, the European Research Council through ERC Advanced Grant SIMCOFE. ZJW and BAB acknowledge the support of MURI-130-6082, ONR-N00014-11-1-0635, NSF CAREER DMR-0952428, NSF- MRSEC DMR-0819860,  Packard Foundation, and a Keck grant. XD is supported by the National Natural Science Foundation of China, the 973 program of China (No. 2011CBA00108 and No. 2013CB921700), and the "Strategic Priority Research Program (B)" of the Chinese Academy of Sciences (No. XDB07020100). 
 
\appendix

\section{Classification of Weyl points}
Type II Weyl points occur when type I Weyl nodes are tilted enough -- along some specific direction -- that a Lifschitz transition occurs and the system acquires finite density of states at the Weyl node.  Although several possibilities of small tilting of the Weyl cone were discussed in some works~\cite{Goerbig-PRB08, Kawarabayashi-IJMP12, Xu-arx14, Chang-PRB15, Trescher-PRB15}, none of these works pointed out the Lifschitz transition that gives rise to a fundamentally new type of Weyl fermion, with new thermodynamic properties. Given the existence of at least two types of Weyl fermions, we can ask if any other types of Weyl points can exist. In the present section we show that a mathematical theorem states that only two types of Weyl points are possible, and hence our  classification is complete. 

The most general linear in ${\bf k}$ Hamiltonian of a Weyl point is
\begin{equation}
H({\bf k})={\bf v}\cdot{\bf k} +\sum_{i,j=1}^3 k_i A_{ij} \sigma_j
\end{equation}
where $A$ is a 3$\times$3 matrix and $\sigma$'s are the Pauli matrices. The Fermi surface for the chemical potential $\mu$ can be obtained by solving the equations $\epsilon_\pm({\bf k})=\mu$, where 
\begin{equation}
\epsilon_\pm({\bf k})={\bf v }\cdot{\bf k}\pm\sqrt{\sum_{i,j} k_i [AA^T]_{ij} k_j}
\label{eq2}
\end{equation}
are the two eigenvalues of the Hamiltonian. By squaring the Eq.~\ref{eq2} the Fermi surface can be shown to be satisfied on the quadric surface described by the following equation
\begin{equation}
\sum_{i,j} k_i([AA^T]_{ij}-v_i v_j)k_j+2\mu {\bf  v}\cdot {\bf k}-\mu^2=0
\end{equation}
Classification of all possible quadric surfaces is known~\cite{Hilbert-book}. Applying this classification to the equation above one can show that there only two possible types of Weyl points: type-I with a closed point-like Fermi surface, and type-II with an open Fermi surface, as described in this paper. The classification~\cite{Hilbert-book} also gives rise to different kinds of nodal lines and surfaces, which will be the subject of further study~\cite{genroute}. 

\section{Transport signatures of type-II Weyl semimetals}
Here we show that the transport properties of the two different types of Weyl points are very distinct. In particular, the type-II Weyl points give rise to a new kind of chiral anomaly. Here we report the main conclusions, and a more detailed explanation will be presented elsewhere~\cite{LLs}. We also discuss another difference between the two types of Weyl points, which appears in the density of states, leading to very different thermodynamic properties. 

\subsection{Novel chiral anomaly in type-II Weyl semimetals}
Let us consider the Hamiltonian 
\begin{equation}
H({\bf k})=Ck_z+{\bf k}\cdot {\bf \sigma}
\label{hamex}
\end{equation}
which realizes a Weyl point with a $+1$ Chern number. The type of this Weyl point depends on the value of the parameter $C$, being of type-II for $|C|>1$ and of type-I otherwise. This Hamiltonian is a much simplified version for that of a type-II Weyl, but has the advantage of being analytically tractable in some limits and will be used for qualitative purposes. 

If a magnetic field of magnitude $B$ is applied along the $\hat{z}$-direction the spectrum of the Landau levels is given by
\begin{equation}
E_n^{\pm}=Ck_z\pm\sqrt{k_z^2+\frac{2n}{\ell^2}}
\end{equation}
where $\ell=(eB/c)^{-\frac{1}{2}}$ is the magnetic lengthscale, $e$ is the charge of electron, $c$ is the speed of light, and $n>0$. The 0-th Landau level  has the energy 
\begin{equation}
E_0=(C+1)k_z
\end{equation}
and is unpaired (chiral). The sign of the electron velocity associated with the 0-th Landau level is given by the sign of $(C+1)$, as explained below. 

If the electric field is applied in the same direction as the magnetic field, chiral electrons start flowing in the $\hat{z}$-direction. Since Weyl points come in pairs of opposite Chern numbers, the low-energy spectrum of a Weyl semimetal will also host an antichiral 0-th Landau level at the position of the opposite Chern number Weyl point. The current then appears flowing from one Weyl point to another~\cite{Volovik-JETPL86, Volovik-book, Volovik-Nature97, Volovik-PhysicaB98, Zyuzin-PRB12, Son-PRB13, Liu-PRB13, Kharzeev-PRB13, Hosur-crp13, Volovik-JETPL14, Burkov-PRL14}. This is a realization of the Adler-Bell-Jackiw chiral anomaly~\cite{Adler-PR69, Bell-Cimento69} of the quantum field theory in the lattice system~\cite{Nielsen-PLB83}. Observation of this effect was reported recently in Dirac semimetals and in the TaAs type-I Weyl semimetal family~\cite{Zhang-arx15, Huang-arx15b, Xiong-arx15}. For the type-II Weyl semimetal similar behavior is expected for the field applied in the direction of dominating kinetic term, when the spurious large trivial almost compensated Fermi surfaces that exist in the material are discounted and the Weyl contribution is singled out. 

\begin{figure}
\begin{center}
\includegraphics[width=\columnwidth]{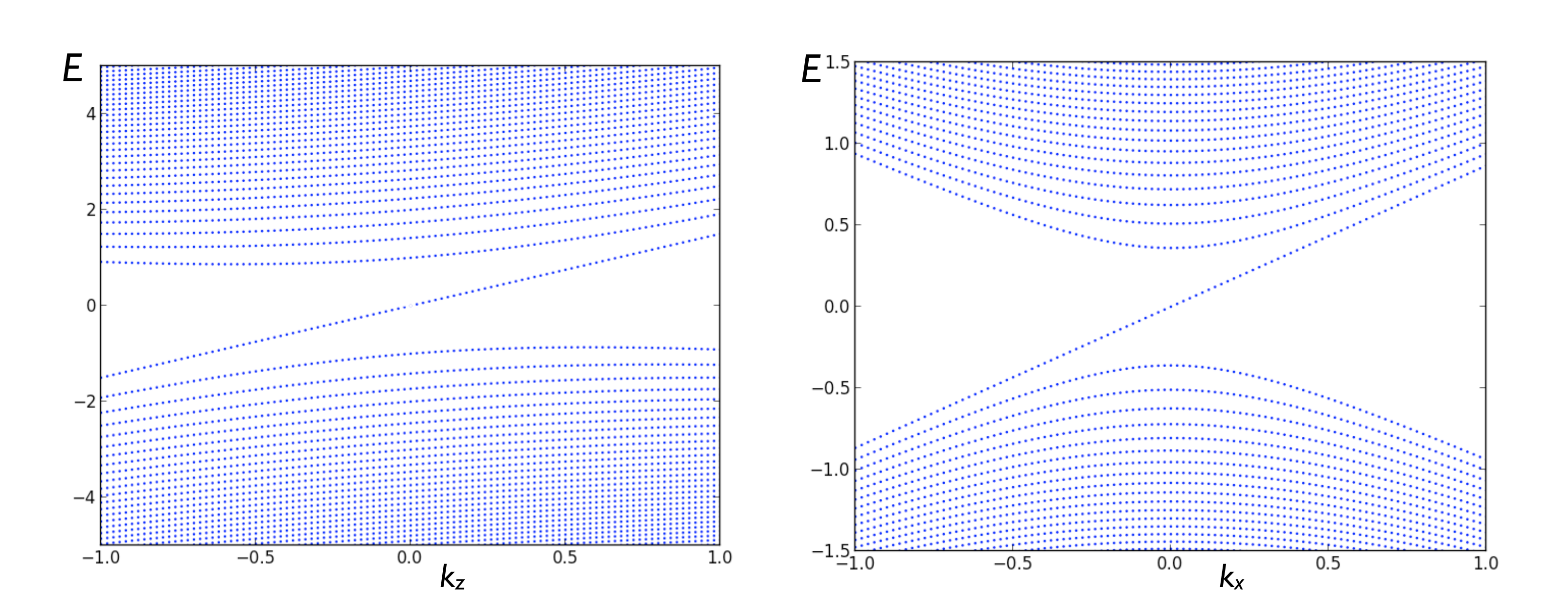}
\end{center}
\caption{Landau level spectrum for the type-I Weyl point of Eq.~\ref{hamex} with $C=0.5$. Left panel: the magnetic field is applied along the $\hat{z}$-direction. Right panel: the magnetic field is applied in the $-\hat{x}$ direction. The chiral 0-th Landau level is present in both cases.}
\label{fig:LL1}
\end{figure}

\begin{figure}
\begin{center}
\includegraphics[width=\columnwidth]{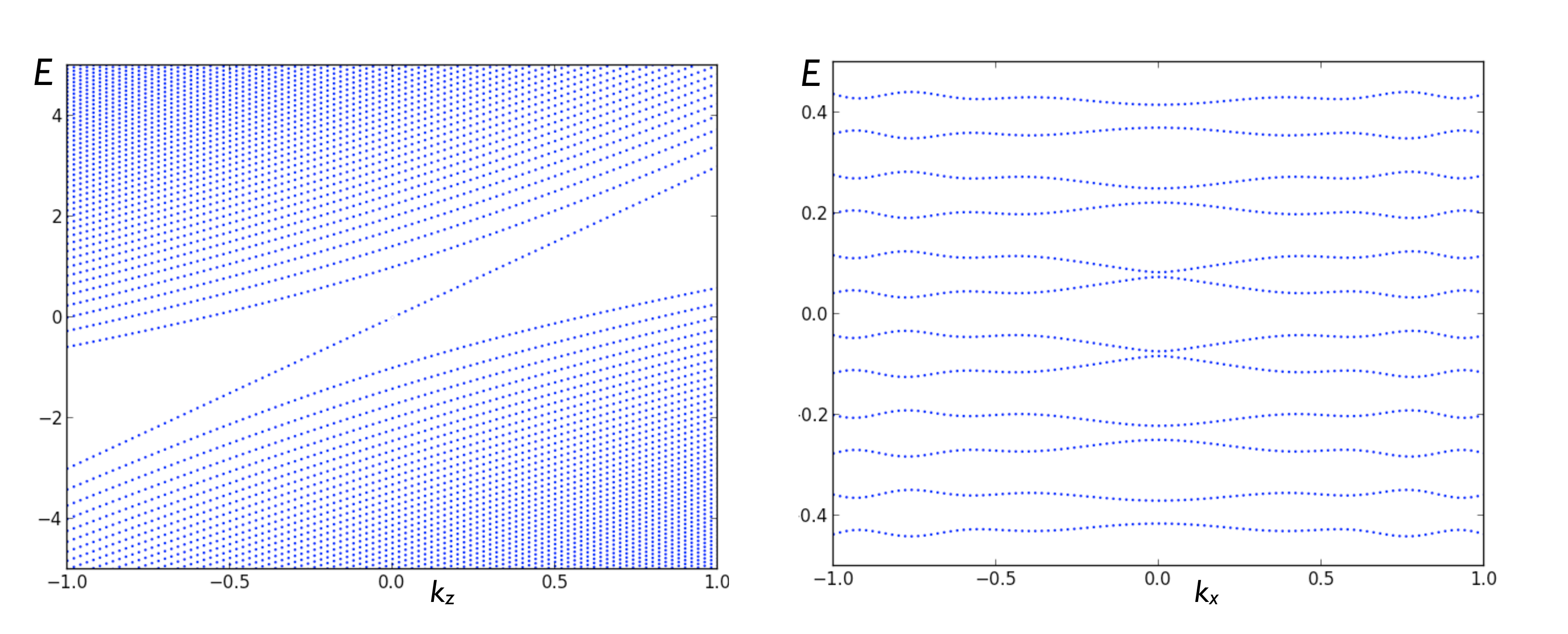}
\end{center}
\caption{Landau level spectrum for the type-II Weyl point of Eq.~\ref{hamex} with $C=2$. Left panel: the magnetic field is applied along the $\hat{z}$-direction. The chiral 0-th Landau level is present. Right panel: the magnetic field is applied in the $-\hat{x}$ direction. No chiral Landau levels exist.}
\label{fig:LL2}
\end{figure}

A drastic difference between the two types of Weyl points described by Eq.~\ref{hamex} arises when the magnetic field is applied along the $\hat{x}$-direction, that is in the direction, in which the kinetic term {\it does not} dominate in the case of $|C|>1$. For the type-I the above discussed chiral anomaly still appears. For the type-II case of $|C|>1$, however, the Landau level spectrum becomes gapped and no chiral anomaly can be observed, as illustrated in Figs.~\ref{fig:LL1}-\ref{fig:LL2} ($\frac{\sqrt{2}}{\ell}=0.1$ in all the illustrations). 

This is easily understood in the limit $C \rightarrow \infty$ by solving the $H=C k_z$ hamiltonian in a $B||\hat{x}$ field. In a more general case, a long analytic calculation~\cite{LLs} proves the absence of a chiral mode for $B||\hat{x}$ and $C>1$. In fact, when the field direction is rotated from $\hat{z}$ to $\hat{x}$ the chiral Landau level persists until the angle $\theta$ between the field the $\hat{z}$-axis, becomes such that $\cos^2{\theta}=1/C^2$. For the directions, where $\cos^2{\theta}<1/C^2$ the chiral Landau level disappears.  

This situation is generic in type-II Weyl points. The chiral Landau level appears only when the magnetic field is applied in the direction, along which the kinetic term is dominant. If the field is applied in the other direction -- no chiral Landau level appears. This dependence of the presence or absence of the chiral anomaly on the direction of the applied magnetic field is the main difference between the two types of Weyl semimetals, with immediate consequences for the transport properties. 

\begin{figure}
\begin{center}
\includegraphics[width=\columnwidth]{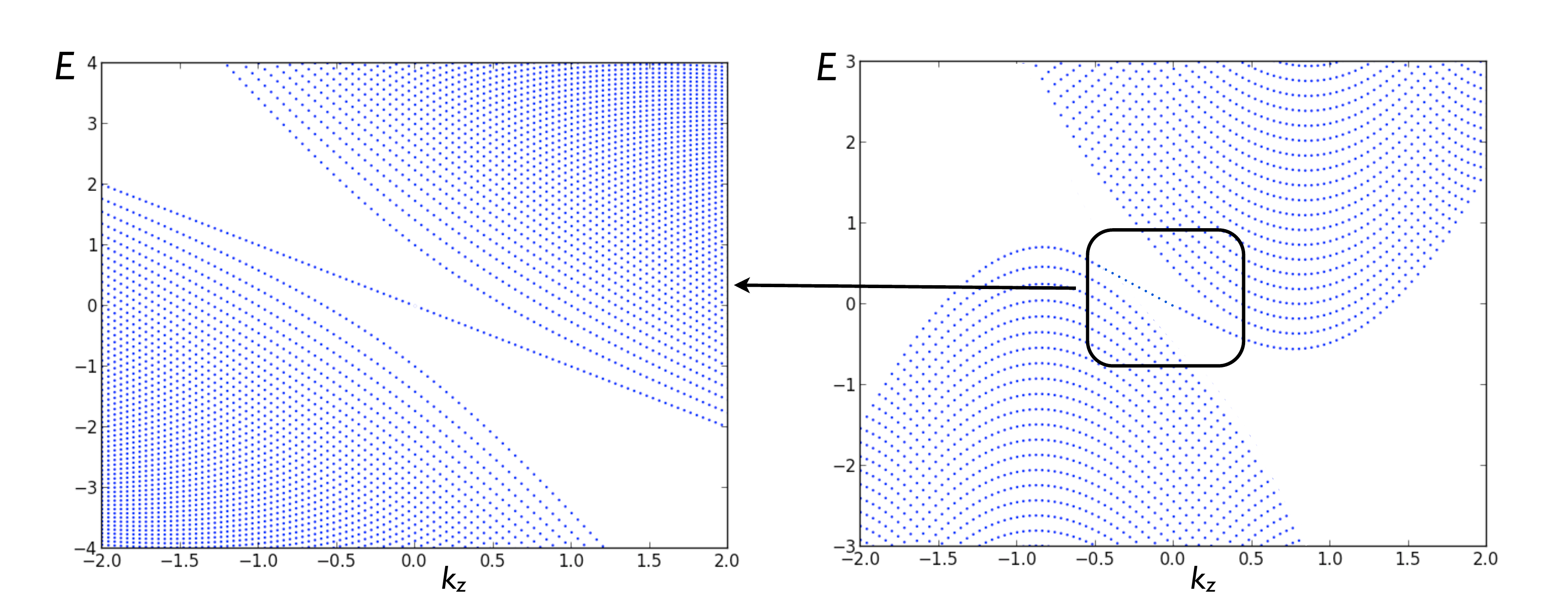}
\end{center}
\caption{Chirality change at small $k_z$. Left panel: Landau level spectrum for the type-II Weyl point of Eq.~\ref{hamex} with $C=-2$. The 0th Landau level appears to be a left mover for a Weyl point with Chern number $+1$. Right panel: Schematic plot based on on a possible realization of the closure of the Fermi surface. The paradox is resolved, when higher-order in $k$ terms are added to the Hamiltonian to close the Fermi surface. These higher order terms render the chirality of the full region to $+1$. The chiral band is a right-mover with the velocity changing twice at small values of $k_z$. The framed region is that, where chirality is changing around small $k_z$.}
\label{fig:LL3}
\end{figure}

There is one more important feature of the type-II Weyl points that does not appear in the type-I case. This is the possibility for the velocity of the chiral Landau level to be {\it locally} in $k$-space different from the chirality dictated by the Chern number of the Weyl node. As an example, consider the Weyl point of the Hamiltonian~\ref{hamex}. The Chern number of this point is positive, however, for $C<-1$ the velocity of the chiral Landau level appearing in the $\hat{z}$-field becomes negative as illustrated in the left panel of Fig.~\ref{fig:LL3}. 

However, this anti-chiral effect is a feature of the linearized Hamiltonian. In a real semimetal higher order terms must exist, responsible for closing the Fermi surface of the type-II Weyl point. These terms will correct the chirality of the Landau level far from the Weyl point as illustrated in the schematic plot in the right panel of Fig.~\ref{fig:LL3}.

\subsection{Thermodynamics of type-II Weyl points}
Another significant distinction of the two types of Weyl semimetals appears in the thermodynamic properties. For the type-I Weyl point, described by the Hamiltonian of Eq.~\ref{hamex} the density of states $g(E)$ behaves like
\begin{equation}
g(E)\propto \frac{E^2}{(1-C^2)^2}
\end{equation} 
For the type-II this behavior changes and the corresponding density of states becomes
\begin{equation}
g(E)\propto \frac{1}{|C|}\left(p_0^2-\frac{E^2(C^2+1)}{(1-C^2)^2}\right)
\end{equation}
The constant term $p_0$ arises due to the presence of unbounded (in the linearized model) electron-hole pockets, and it depends on the cutoff. Due to this term the density of states always remains positive. These formulas were verified versus numerical simulation of the Hamiltonian~\ref{hamex}. 

\begin{figure}
\begin{center}
\includegraphics[width=0.7\columnwidth]{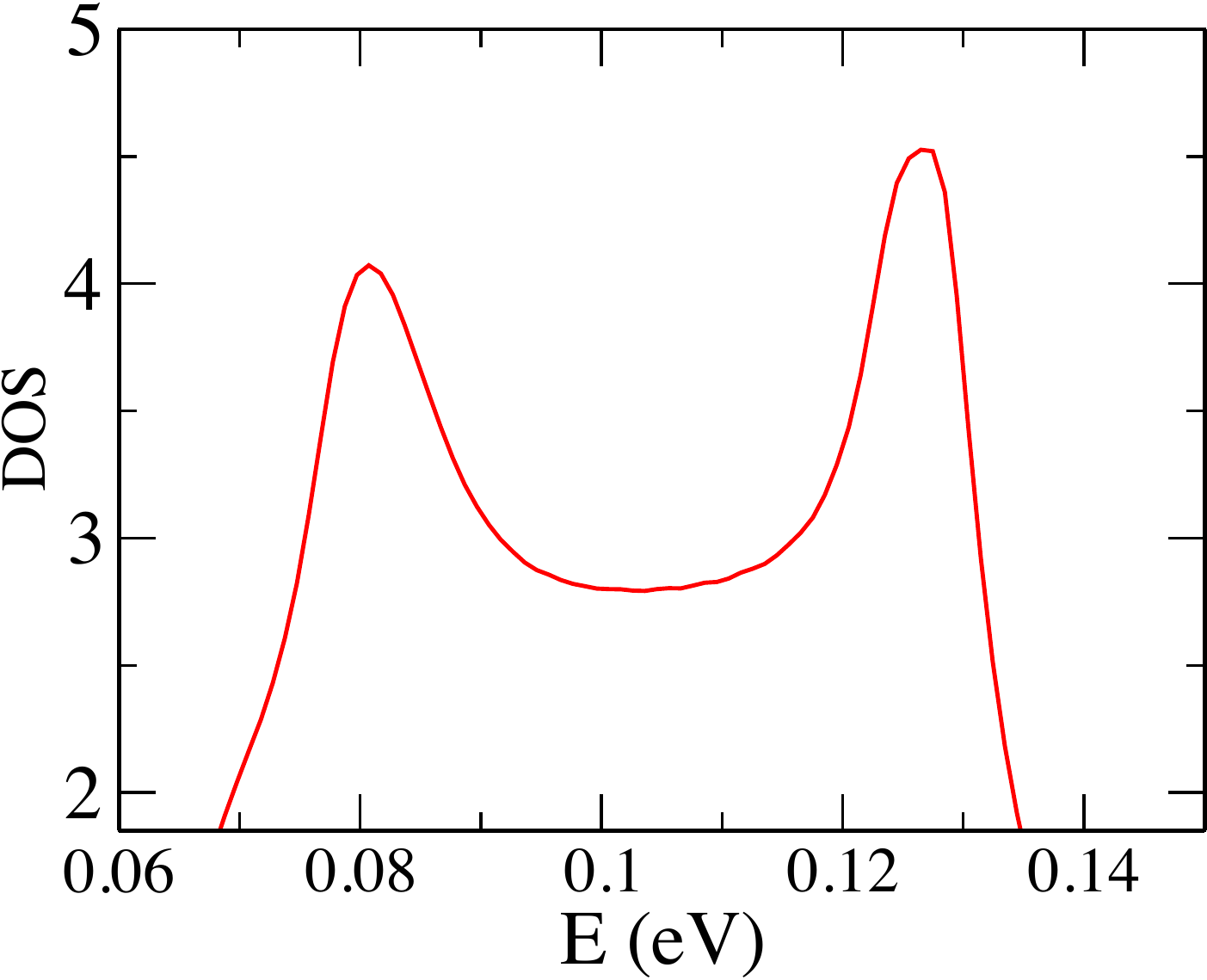}
\end{center}
\caption{The density of states of the pair of Weyl points in WTe$_2$. The contribution of the rest of the Fermi surface is not included.}
\label{fig:dos}
\end{figure}

To illustrate this claim further we plot the density of states that arises due to the pair of Weyl points in WTe$_2$ in Fig.~\ref{fig:dos}. The contribution of the rest of the Fermi surface to the density of states is not included. The two clearly visible parabolic peaks in the density of states are due to the two type-II Weyl points is. The difference in the energy position of the peaks and the energy of Weyl points reported in the main text is due to the difference between the tight-binding model used here with the actual first-principles calculation. The peaks correspond to the energies of Weyl points in the tight-binding model used to generate Fig.~\ref{fig:dos}.
\section{Crystal structure and computation details}
\begin{figure}
\begin{center}
\includegraphics[width=0.7\columnwidth]{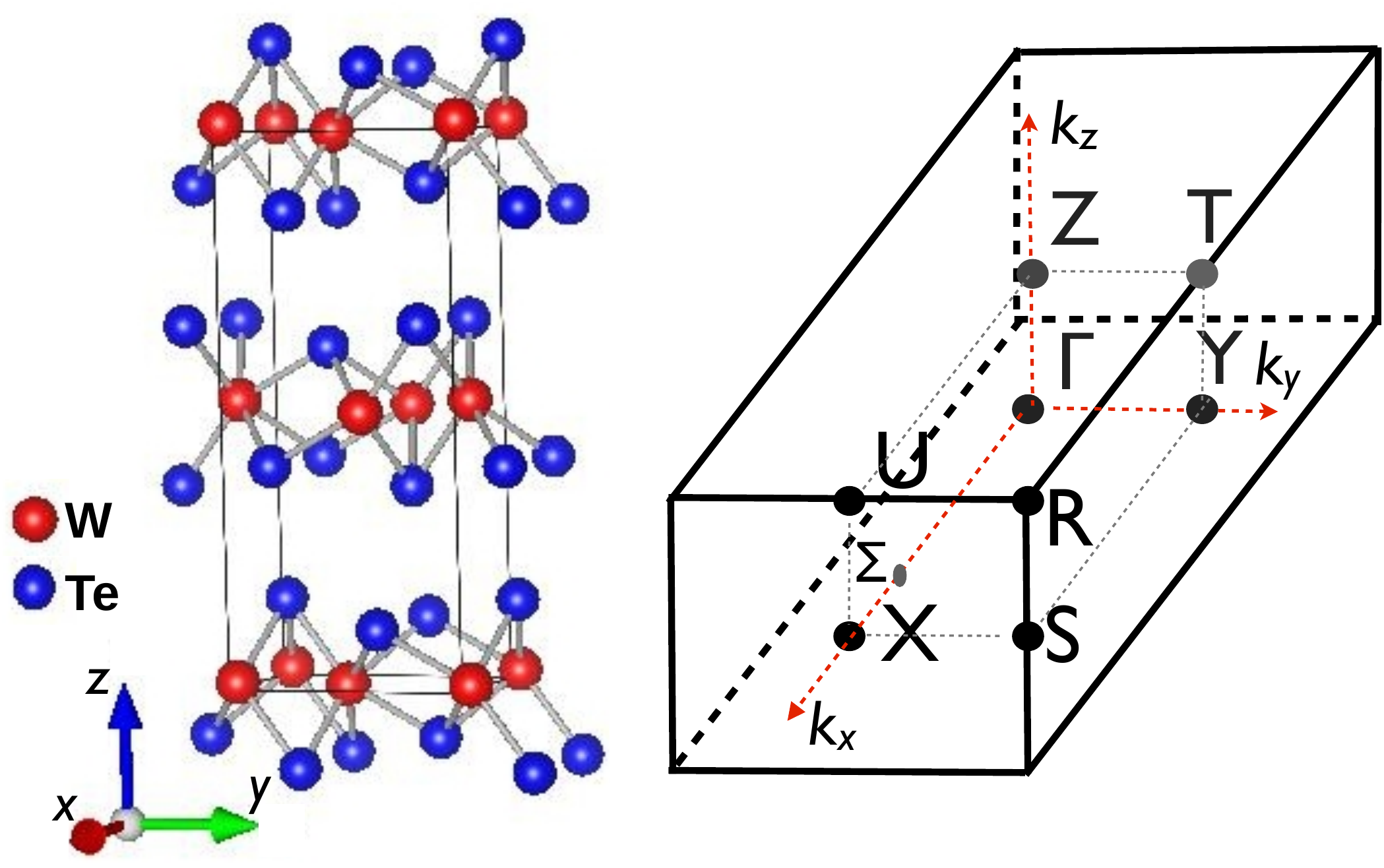}
\end{center}
\caption{Left panel: crystal structure of WTe$_2$. Right panel: Brillouin zone of the rhombohedral unit cell.}
\label{fig:struct}
\end{figure}

The crystal structure of WTe$_2$ and its Brilloin zone are illustrated in Fig.~\ref{fig:struct}.
Several crystal structures are reported in the databases~\cite{brownwte2,wte2latt}. We use the latest one from the work of Ref.~\cite{wte2latt} that was obtained at the lowest temperature. That work explicitly explains minor differences with the earlier works~\cite{brownwte2} by the higher temperatures in those measurements.

For completeness, we list the experimental structural parameters taken from Ref.~\cite{wte2latt} that were used in this work. The lattice constants are $a=3.477\AA$, $b=6.249\AA$ and $c=14.018\AA$, and both W and Te atoms occupy $2a$ Wyckoff positions corresponding to $(0, y, z)$ and $(1/2, -y, z+1/2)$. Values of $x$ and $y$ for this structure are listed in Table~\ref{tab:coord}.
\begin{table}[h]
\begin{center}
\begin{tabular}{|c|c|c|c|c|c|c|}
\hline
 & W(1)& W(2)& Te(1)& Te(2)& Te(3)& Te(4)\\
\hline
 $y$ & 0.60062 & 0.03980 & 0.85761 & 0.64631 & 0.29845 & 0.20722\\
 \hline
$z$ &0.5 & 0.01522 & 0.65525 & 0.11112 & 0.85983 & 0.40387\\
\hline 
\end{tabular}
\caption{Positions of atoms in the unit cell of WTe$_2$ given as coordinates for Wyckoff positions $2a$ of the $Pnm2_1$ space group. Bracketed numbers following the element symbol indicate distinct Wyckoff positions. There are two distinct Wyckoff positions for W and four for Te atoms.} 
 \label{tab:coord}
 \end{center} 
 \end{table}

Band structure calculations (both with and without spin-orbit coupling (SOC)) were performed in VASP~\cite{VASP} {\it ab initio} code using PAW~\cite{PAW1, PAW2} pseudopotentials with $ 6s^2 5d^4$ and $5s^2 5p^4$ valence electron configurations for W and Te correspondingly. The PBE~\cite{PBE} approximation was used. Spin-orbit coupling was implemented in pseudopotentials. The energy cutoff was taken to be $260$eV. Gaussian smearing of width $0.05$eV and a $12\times 10 \times 6$ $\Gamma$-centered $k$-point mesh were used to perform Brillouin zone integrations. The conclusions of the paper were also verified using a more elaborate pseudopotential for W with semicore $p$-states included in the valence, that is with $5p^6 6s^2 5d^4$ valence electron configuration. 

\section{Electronic structure without spin-orbit coupling}

\begin{figure}
\begin{center}
\includegraphics[width=0.7\columnwidth]{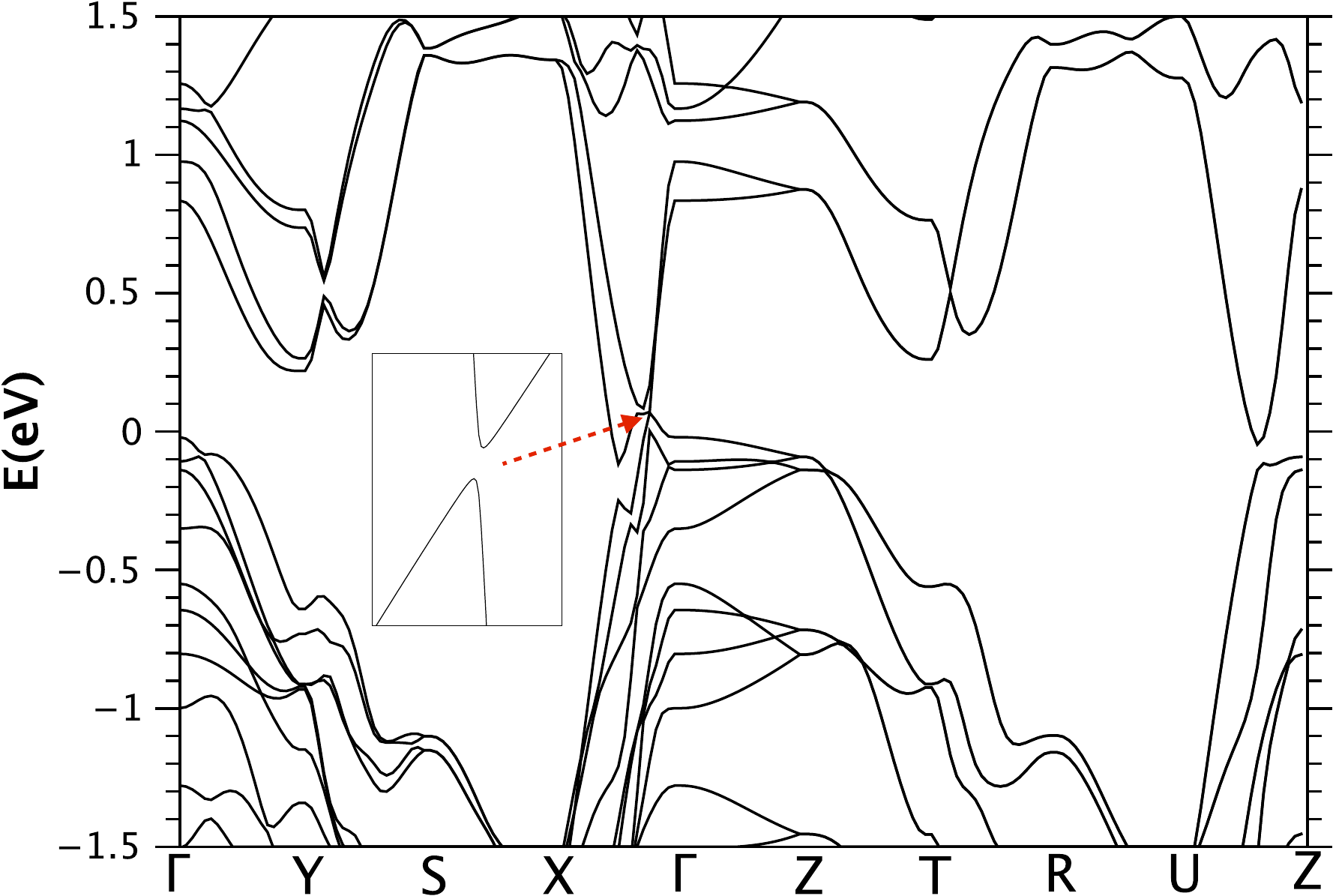}
\end{center}
\caption{Band structure of WTe$_2$ without SOC. $E_{\rm F}$ is set to $0$.}
\label{fig:wte_noso}
\end{figure}

The band structure of WTe$_2$ in the absence of SOC is shown in Fig.~\ref{fig:wte_noso}. To understand the 
nature of crossings in this case we carried out an extensive symmetry analysis that will be reported elsewhere~\cite{genroute}. Here we only state the main conclusions. Using a two band model 
\begin{equation}
H(k_x,k_y,k_z)=\varepsilon(k_x,k_y,k_z)+\sum_{i=x,y,z} d_i(k_x,k_y,k_z)\sigma_i, 
\end{equation}
where $\sigma_i$ are the Pauli matrices, one can show that on the $k_z=0$ plane only two of the three $d_i$ coefficients are linearly independent. The degeneracy occurs if for some $(k_x,k_y)$ both of these coefficients vanish. Thus, there are two independent constraints on two functions of $k_x,k_y$, the codimension is zero, and the solution is in general possible at isolated points in the $k_z=0$ plane. In addition, it can be shown that without SOC no band crossings can generally occur along the $\Gamma-X$ axis. Indeed, {\it ab initio} results show that a small band gap exists in the band structure along $\Gamma-X$, as illustrated in the inset of Fig.~\ref{fig:wte_noso}.

First-principles calculations support our conclusions. Without SOC we find 8 Weyl points (not accounting for spin) in the $k_z=0$ plane. In addition, we find 8 more spinless Weyls at low symmetry points $k_i\neq 0$ for all $i$. The coordinates of Weyl points and their Chern numbers are given in Tab.~\ref{tab:noso}.
\begin{table}[h]
\begin{center}
\begin{tabular}{|c|c|c|c|c|}
\hline
 & $k_x$& $k_y$& $k_z$& $C$\\
\hline
 1 & 0.1054 & 0.0087 & 0 & +1 \\
 \hline
2 & 0.1742 & 0.1200 & 0 & +1 \\
\hline
3 & 0.14992   & 0.07845 & 0.2329936 & -1\\
\hline
4 & 0.14992 & 0.07845 & -0.2329936 & -1\\ 
\hline 
\end{tabular}
\caption{Coordinates and Chern numbers of 4 out of 16 Weyl points appearing without SOC. The other 12 points are related by the ones listed by reflections being located at $(-k_x,k_y,k_z)$, $(k_x, -k_y,k_z)$ with Chern numbers $-C$ and at $(-k_x,-k_y,k_z)$ with Chern number $C$. Points at $k_z\neq 0$ are symmetric about the $xy$-plane due to the compound symmetry formed by time reversal and two-fold rotation.} 
 \label{tab:noso}
 \end{center} 
 \end{table}

Now, when spin is taken into consideration but SOC is still neglected, each Weyl point becomes doubled, with doubled chirality due to $\mathrm{SU}(2)$ symmetry. When SOC is gradually increased the Weyl points split and start moving in the Brillouin zone. For a material with weak SOC, one can expect this motion to be the only effect of SOC, so that the Weyl points do not vanish in general. In WTe$_2$, however, SOC is large. To see its effect on the structure of Weyl points in the BZ, we carried out a smooth interpolation between no SOC and full SOC band structures. In the process, other Weyl points appear and disappear (by merging points with opposite chiralities), so that the final arrangement of Weyl points reported in the main text cannot be understood in terms of rearranging the Weyl points that are present in the absence of SOC.

\section{Electronic structure with spin-orbit coupling}

\begin{figure}
\begin{center}
\includegraphics[width=0.7\columnwidth]{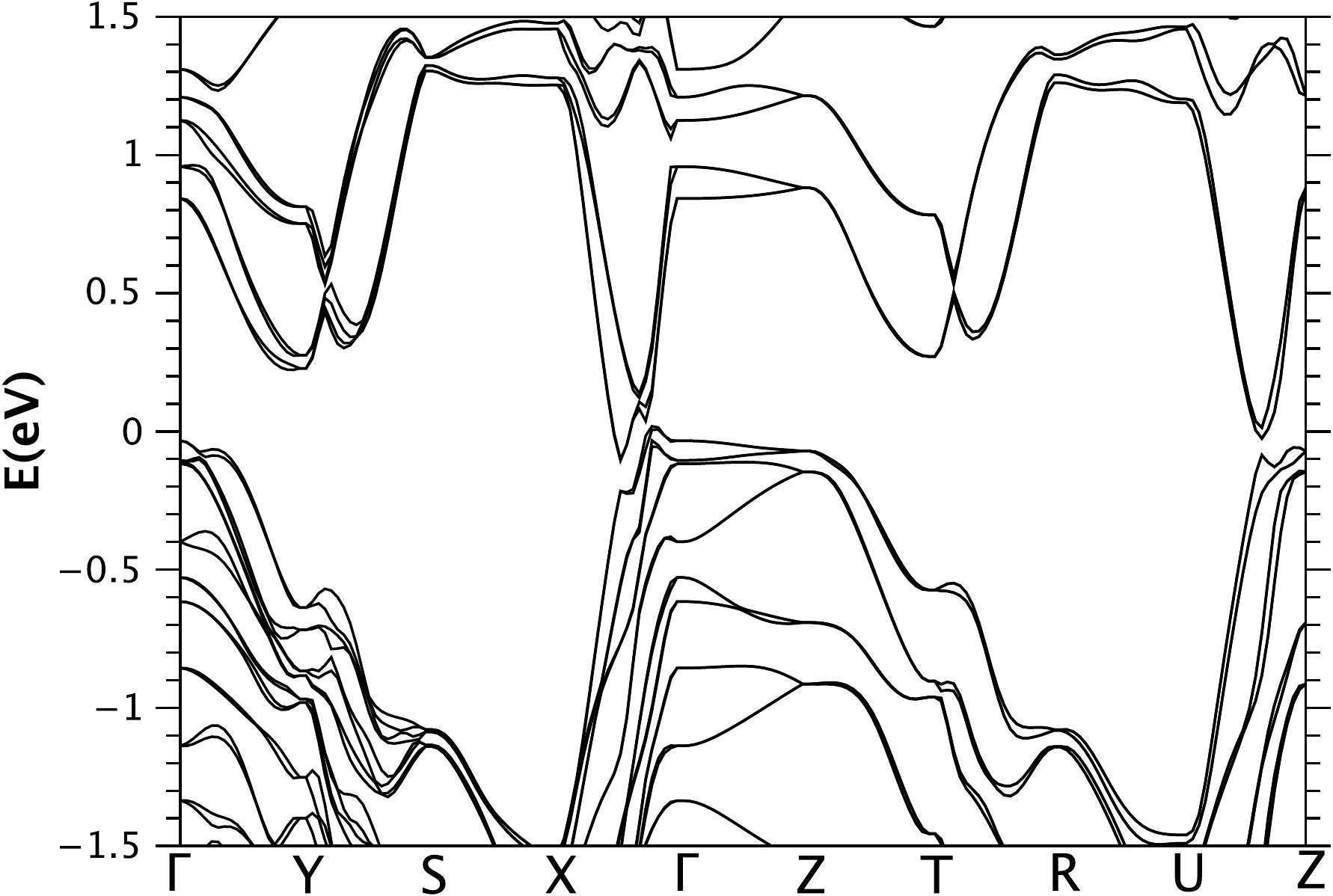}
\end{center}
\caption{Band structure of WTe$_2$ with SOC taken into account. $E_{\rm F}$ is set to $0$.}
\label{fig:wte_so}
\end{figure}

The fully relativistic band structure is illustrated in Fig.~\ref{fig:wte_so}. Symmetry considerations
indicate the possibility to have locally stable point-like degeneracies in the $k_z=0$ plane, where a little group formed by a combination of a two-fold rotation and time-reversal exists~\cite{genroute}. As will be explained elsewhere in detail, this suggests that double degeneracies can occur at isolated points in the $k_z=0$ plane.

In accord with this argument, eight Weyl points (listed in Tab.~\ref{tab:soc}), described in the main 
\begin{table}[h]
\begin{center}
\begin{tabular}{|c|c|c|c|c|}
\hline
 & $k_x$& $k_y$& $k_z$& $C$\\
\hline
1 & 0.12184 & 0.03825 & 0 & +1 \\
 \hline
2 & 0.12141 & 0.0454 & 0 & -1 \\
\hline
\end{tabular}
\caption{Coordinates and Chern numbers of 2 out of 8 Weyl points appearing with SOC. The other 6 points are related by the ones listed by reflections, being located at $(-k_x,k_y,k_z)$, $(k_x, -k_y,k_z)$ with Chern numbers $-C$.} 
 \label{tab:soc}
\end{center}  
 \end{table}
text, are indeed located in the $k_z=0$ plane. A general form of the Hamiltonian around a Weyl point in the $k_z=0$ plane can be constructed using the following two dimensional representation for the combination of $C_2$ and time-reversal
\begin{equation}
C_{2T}=-i\sigma_z K
\end{equation}
where $K$ is complex conjugation. This form of the symmetry can be shown to be consistent with the representations of other symmetries in this non-symmorphic space group. 

\begin{figure}
\begin{center}
\includegraphics[width=0.7\columnwidth]{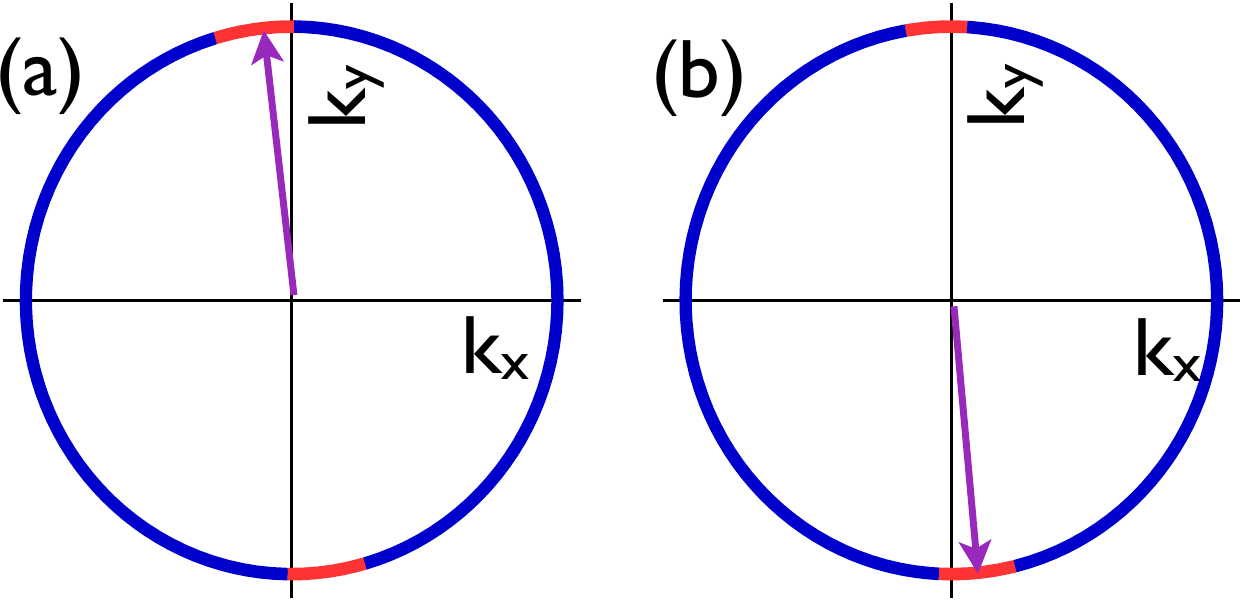}
\end{center}
\caption{The ratio $R$ of Eq.~\ref{ratio} for the circle of radius $10^{-3}$ drawn around the two Weyl points. Red color corresponds to $R>1$, while blue is $R<1$. The arrows show the direction from one of the two points to the other. Panel (a): Weyl point at $(0.12184,0.03825,0)$. Panel (b): Weyl point at $(0.12141, 0.0454, 0)$.}
\label{fig:kin}
\end{figure}

Keeping only terms linear in $k$ and subjecting the Hamiltonian to this symmetry we find the general form of the Hamiltonian around a Weyl point in $k_z=0$ with SOC
\begin{equation}
H({\bf k})=\epsilon({\bf k})+ (ak_x+ck_y)\sigma_y +(bk_x+dk_y)\sigma_z +e k_z \sigma_x
\label{hamk}
\end{equation}
where 
\begin{equation}
\epsilon({\bf k})=Ak_x+Bk_y
\end{equation}
For the Weyl point to be of type II, the kinetic part of the energy should dominate over the potential one in at least some direction in $k$-space. To find such directions for the points in question, we plotted the ratio
\begin{equation}
R=\frac{(Ak_x+Bk_y)^2}{e^2k_z^2+(ak_x+ck_y)^2+(bk_x+dk_y)^2}
\label{ratio}
\end{equation}
on the circles defined by $k^2_x+k^2_y=10^{-6}$, where $k_x$ and $k_y$ are taken in reduced coordinates, and the constants are explained in the main text. The results are illustrated in Fig.~\ref{fig:kin}. The region, where $R>1$, meaning that the kinetic energy dominates and the Weyl points are of type II, is shown in red.

Finally, we fitted the Hamiltonian of Eq.~\ref{hamk} to the first-principles band structure around the Weyl points to get the values of its parameters. They are provided in Tab.~\ref{tab:param}. 
\begin{table}[h]
\begin{center}
\begin{tabular}{|c|c|c|c|c|c|c|c|}
\hline
point & $A$& $B$& $a$& $b$ & $c$ & $d$ & $e$\\
\hline
1 & -2.739 & 0.612 & 0.987 & 1.107& 0.0 & 0.270& 0.184\\
 \hline
2 & 1.204 & 0.686 & -1.159 & 1.046& 0.0 & 0.055& 0.237\\
\hline
\end{tabular}
\caption{Parameter values for the Hamiltonian of Eq.~\ref{hamk} around the two Weyl ponts given in [eV\AA]. Point 1 refers to the Weyl at $(0.12184,0.03825,0)$, while point 2 is $(0.1241,0.0454,0)$.} 
 \label{tab:param} 
\end{center}
 \end{table}
\section{Weyl Nodes and Topological Indices}

To rigorously establish the degeneracies of WPs, we examine the structure of the Berry curvature ${\cal F}({\bf k})$ of Bloch states around the gapless points.  Since a WP represents a monopole in Berry curvature, the flux of ${\cal F}$ through a surface enclosing it in $k$-space is quantized to the total topological charge enclosed, and the FS of WP1, being a closed surface, has non-zero Chern number. In case of WP2, however, the FS is open, and  cannot be used to compute its topological charge. Instead, we integrate the Berry curvature of $N$ bands, where $N$ is the number of electrons per unit cell. A surface, on which these $N$ bands have an energy gap to the higher energy bands and which encloses the WP2 can always be found and is used to compute the topological charge of a WP2, as illustrated below.    

\begin{figure}
\begin{center}
\includegraphics[width=\columnwidth]{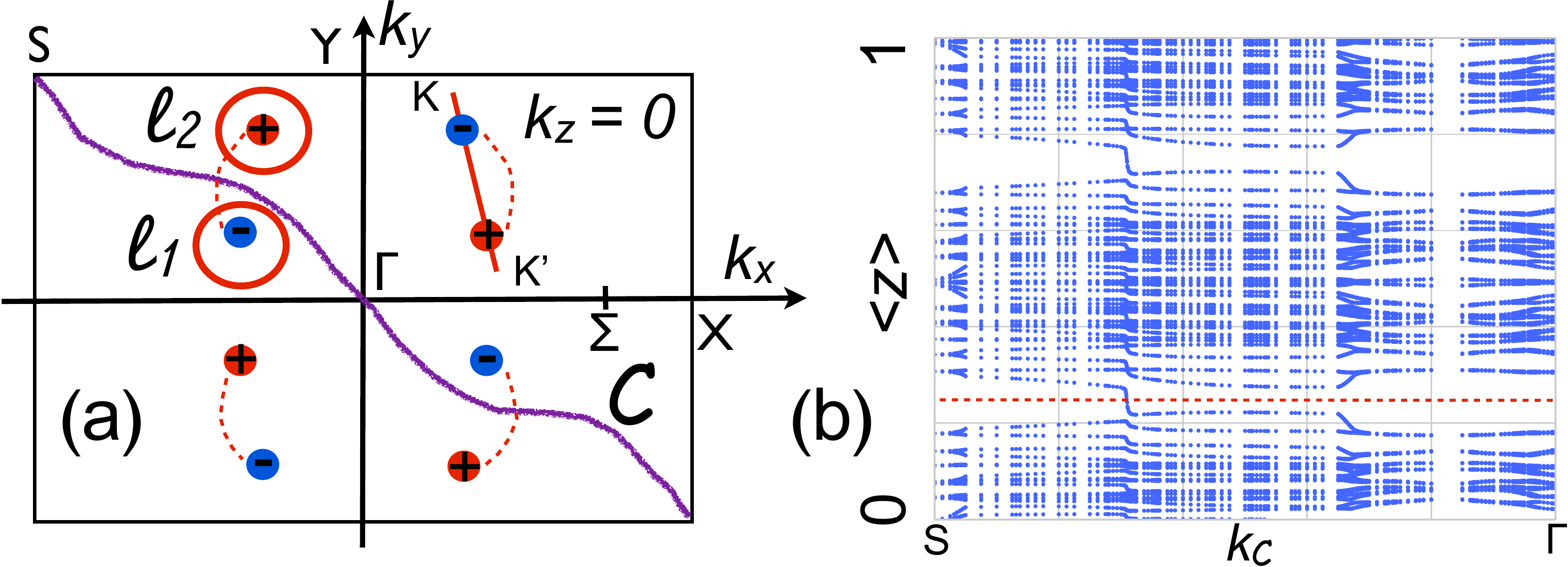}
\end{center}
\caption{Weyl points of WTe$_2$. Panel (a): schematic illustration of the BZ cross-section at $k_z=0$ and Weyl points in it, their chirality marked with red and blue color, corresponding to $C=+1$ and $C=-1$. The red line corresponds to the BZ segment shown in Fig.~\ref{fig:bstrs}(c). Panel (b): Wannier charge centers (Wilson loop) on the plane, shown schematically as contour ${\cal C}$ in panel (a). It corresponds to a non-trivial $\mathbb{Z}_2$ invariant.}
\label{fig:cherns}
\end{figure}

For WTe$_2$ these surfaces were taken to be spheres enclosing the Weyl points; see $\ell_1$ and $\ell_2$ in Fig.~\ref{fig:cherns}(a). For the sphere $\ell_1$ the Chern number is found to be $C=-1$,  proving the existence of a WP of negative chirality inside it. The sphere $\ell_2$, enclosing the second WP, is found to have  $C=+1$. This calculation combined with the use of symmetries maps out the structure of the eight WPs in the BZ of WTe$_2$. 
They are sketched in Fig.~\ref{fig:cherns}(a).

To understand which pairs of Weyl nodes are connected by the Berry flux, thus elucidating the structure of the surface Fermi arcs, several topological invariants are computed from first-principles calculations.  TR allows the computation of the $\mathbb{Z}_2$-invariant~\cite{Kane-PRL05-b} on TR-symmetric planes in the BZ, defined by $k_i=0,\pi$, where $i=x,y,z$. One such plane, the $k_z=0$ one, hosts WPs, and does not have a well-defined invariant. For the other five planes, where the $N=72$ bands are gapped from the band $N+1$, the topological invariant is computed using a hybrid Wannier centers technique~\cite{Soluyanov-PRB11-b} and found to be trivial.  The corresponding Wilson loops~\cite{Yu-PRB11} are gapped.

A non-trivial TR $\mathbb{Z}_2$ topological invariant can, however, be defined for WTe$_2$. It describes the motion of Wannier charge centers on a curved surface that crosses neighboring WPs, as shown by contour ${\cal C}$ in Fig.~\ref{fig:cherns}(a) where we plot its projection onto the $(k_x, k_y)$-plane - the surface extends through the BZ in the $k_z$-direction. If the curve passes through a point $(k_x,k_y)$ then it passes through its TR-image $(-k_x,-k_y)$ as well and the surface satisfies the conditions of the $\mathbb{Z}_2$-pump ~\cite{Fu-PRB06}. We calculate $\mathbb{Z}_2$ invariant on this surface  directly from first-principles calculations using the Z2Pack software ~\cite{Z2pack}.  Fig.~\ref{fig:cherns}(d) shows that pumping in this plane occurs between two TR invariant momenta: $k_x=k_y=-\pi$ ($S$) and $k_x=k_y=0$ ($\Gamma$). Thus, the plane $({\cal C},k_z)$ exhibits a quantum spin Hall effect. If an open surface in the $z$-direction is introduced, helical surface states appear along the curve ${\cal C}$ in the surface BZ, becoming part of the Fermi arcs.  As the $k_{x,y}=0, k_z$ planes have gapped Wilson loops, we conclude that the Berry flux connects pairs of neighboring WPs, as illustrated in Fig.~\ref{fig:cherns}(a). This calculation gives another  proof that WPs exist in WTe$_2$, resolving their connectivity and the location of Fermi arcs.   

\subsection{Computation of chiralities of type-II Weyl points}

The Fermi surface of a type-II Weyl point is open, so it cannot be used for integrating the Berry curvature in chirality computation. Instead, the Berry curvature of $N$ bands, where $N$ is the number of electrons per unit cell, should be integrated over a surface, on which these $N$ bands are separated by an energy gap from the higher energy ones. Since the crossing of the $N$th and $N+1$th bands occurs at a point (Weyl point), such a surface enclosing the type-II Weyl point always exists.  The calculation becomes equivalent to the usual integration over the occupied bands, if one allows for a $k$-dependent chemical potential, located in between  the $N$-th and $N+1$-th bands at each $k$-point on the surface.

\begin{figure}
\begin{center}
\includegraphics[width=\columnwidth]{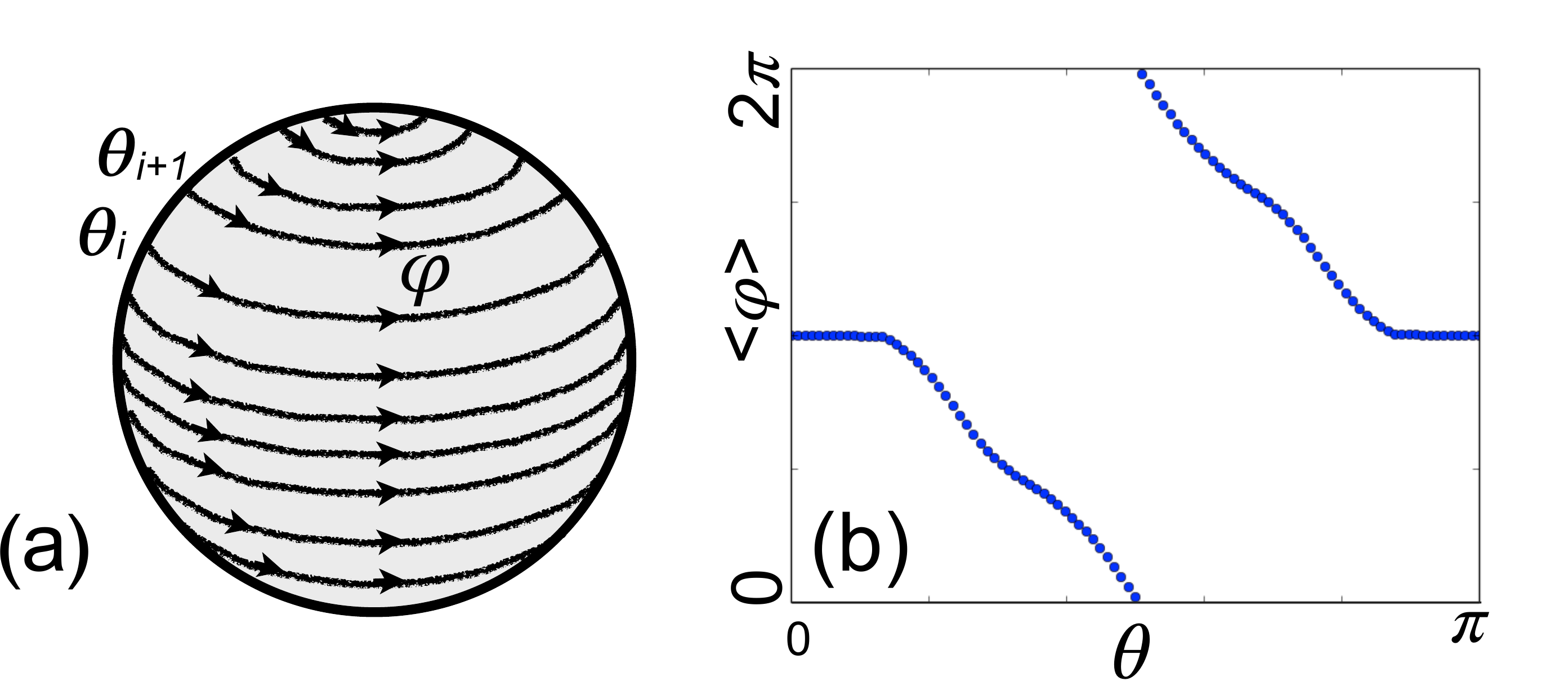}
\end{center}
\caption{Panel (a): schematic illustration of the integration paths used to calculate topological charges of Weyl points. Panel (b): motion of the center of charge around the sphere is shown schematically for $\ell_1$ in panel (a) of Fig.~3 of the main text. The Chern number of the enclosed Weyl point is equal to $-1$.}
\label{fig:chern}
\end{figure}

Following this prescription and using a Wannier function-based tight-binding model~\cite{Souza-PRB01, Wannier90-2} described below, Bloch states were calculated on the spheres around the gapless points. The Hamiltonian remained gapped in the above sense ($N=72$) everywhere on these spheres, schematically illustrated as $\ell_{1,2}$ in Fig.~3(a) of the main text. The corresponding flux of Berry curvature was computed by discretizing a closed sphere, parametrized by angles $\theta$ and $\phi$, into 1D-loops, as shown in Fig.~\ref{fig:chern}(a). When traversed around a loop, the occupied Bloch states accumulate a Berry phase, the trace of which is computed for each of these loops $\theta_i$ using the methods of Refs.~\cite{Soluyanov-PRB11-a, Soluyanov-PRB11-b, Yu-PRB11}. 

These Berry phases correspond to the average position of charge~\cite{King-Smith-PRB93}, associated with the bands below the gap, on the loop, $\langle \varphi \rangle(\theta_i)$. Since 1D loops cover a closed surface,  the center of charge $\langle \varphi \rangle$ can only shift by an integer multiple of $2\pi$ when $\theta$ varies from $0$ to $\pi$. This multiple is equal to the monopole charge enclosed and gives the chirality of the WP enclosed. Similar considerations are used when computing Chern numbers of insulators, with the only difference that the BZ in that case is a torus, rather than a sphere. The result of such a calculation is equivalent to taking the surface integral of the Berry curvature over a closed surface. A more detailed account of this type of calculations, along with the rigorous derivation can be found in Ref.~\cite{Z2Pack}. 

The results obtained for WTe$_2$ is shown in  Fig.~\ref{fig:chern}(b) for the sphere $\ell_{1}$ of Fig.~3(a) of the main text. The charge center shifts downwards, corresponding to Chern number $C=-1$, thus proving the existance of a WP of negative chirality inside $\ell_1$. For the sphere $\ell_2$, enclosing the second Weyl point, the Chern number is $C=+1$, and hence the chirality of this point is positive. 

\section{Strain effects in WTe$_2$}

As mentioned in the main text, the hydrostatic compressive strain (applied pressure) of 2\% increases the separation of the Weyl nodes from 0.7\% to 4\% of the reciprocal lattice constant separation of the nearest Weyl points. Other strains were also studied. Here we present the summary of the results obtained.

\begin{figure}
\begin{center}
\includegraphics[width=\columnwidth]{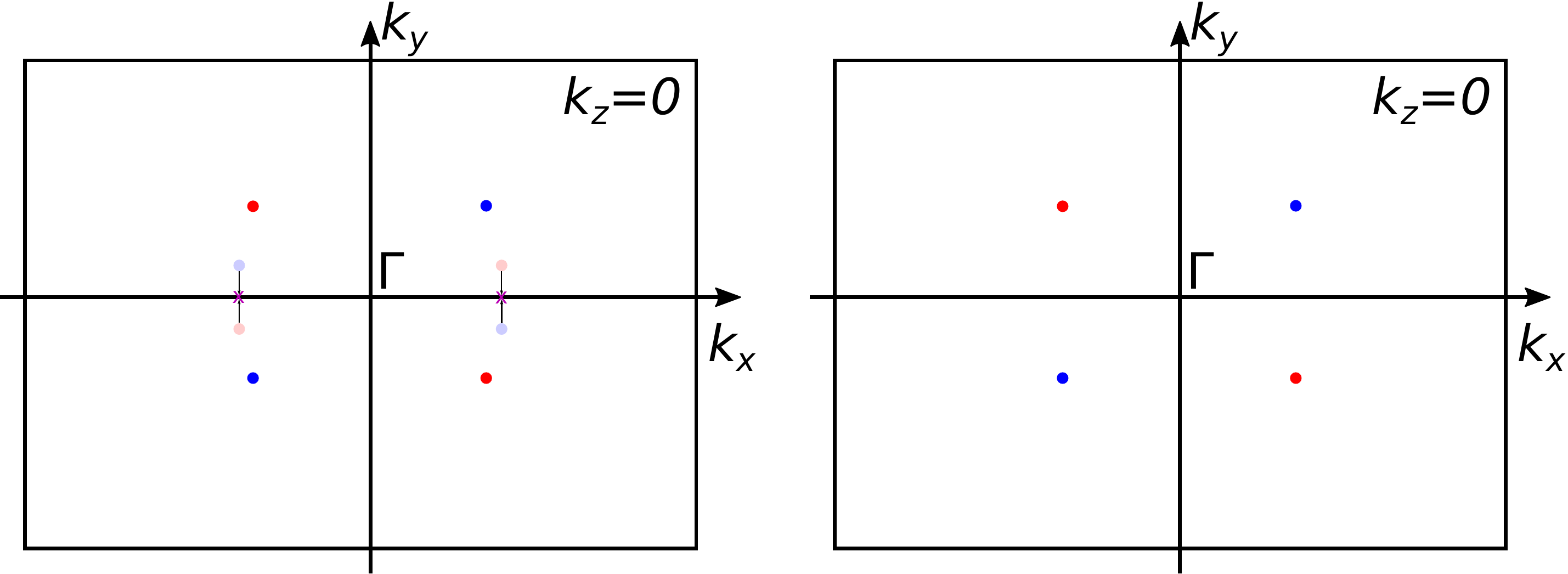}
\end{center}
\caption{Schematic illustration of the effect of uniaxial compressive strain along the $\hat{x}$-direction. Left panel: at the strain of $\approx$2\% four Weyl points meet in pairs on the mirror plane and annihilate. Right panel: only four Weyl points are left in the $k_z=0$ plane.}
\label{fig:strain4}
\end{figure}

Compressive uniaxial strain along the $x$-direction also increases the separation between the nearest Weyl points. At 1\% strain the separation is 2.2\% of $|G_2|$. Further increase of this strain makes one of the two points hit the mirror plane (at $\approx$2\% strain), where it is annihilated with its mirror image of opposite chirality, and thus only four Weyl points are left in the $k_z=0$ plane, all of them being of type-II. This situation, illustrated if Fig.~\ref{fig:strain4}, possibly realizes the simplest situation in time-reversal symmetric Weyl semimetals with the minimal number of Weyl points.

\begin{figure}
\begin{center}
\includegraphics[width=\columnwidth]{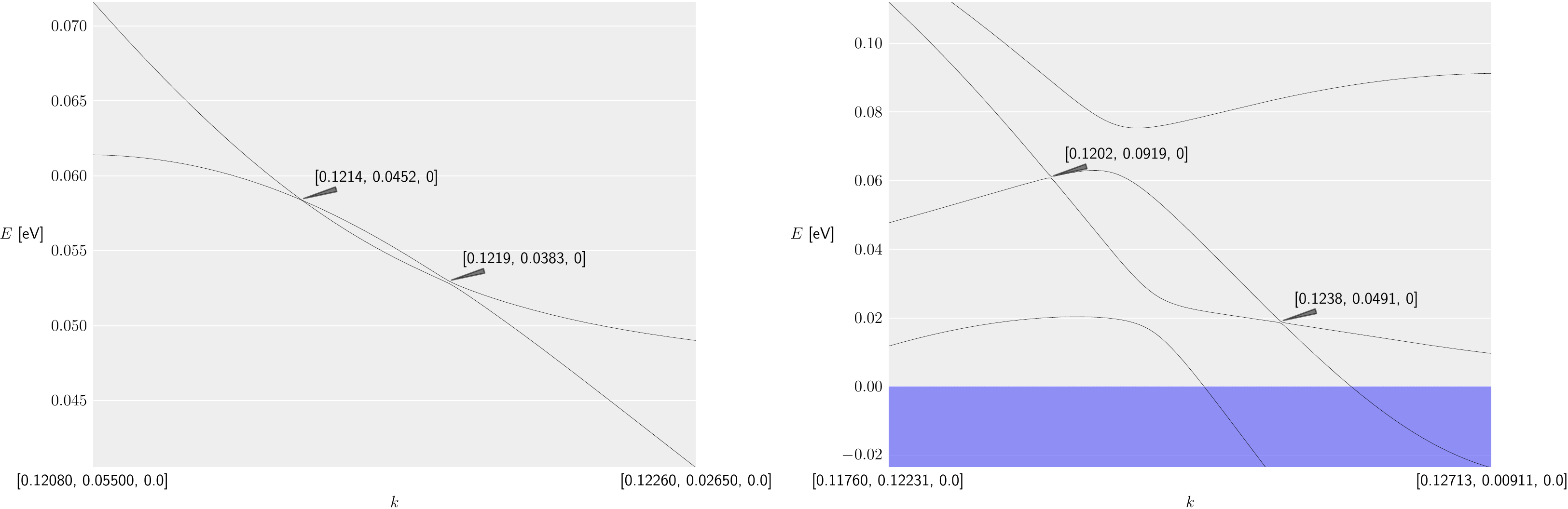}
\end{center}
\caption{Illustration of type-II to type-I Weyl point transition under the compressive uniaxial strain in the $\hat{z}$-direction. Left panel: no strain. All points are of type-II. Right panel: 2\% strain. The points at higher energy were checked in 3D $k$-space to be of type-I. The points move further apart in energy, one of them appearing very close to $E_{\mathrm{F}}$.}
\label{fig:2to1}
\end{figure}

The situation changes for the uniaxial compressive strain in the $z$-direction. While the the neighboring Weyls still move further from each other (4.3\% of $|G_2|$ for a 2\% strain), they both move further away from the mirror plane, so the scenario of only four Weyl points left is not realized in this case. However, we find that this strain drives a phase transition from type-II to type-I Weyl points for four out of eight points. Those, that are furthest from the mirror plane become type-I at $\approx$2\% strain as illustrated in Fig.~\ref{fig:2to1}.

Stretching the crystal in both $x$ and $z$ directions leads to the mutual annihilation of the neighboring Weyl points and drives the transition from a topological semimetal with eight Weyl points to the trivial semimetal with no Weyl points at all. For the uniaxial $y$-strain the situation is reversed. Stretching moves the neighboring Weyl points away from each other (2.5\% $|G_2|$ for a 2\% stretch) and for reasonable strains the number of Weyl points does not change, while compression in this direction leads to annihilation of all points.

We also considered a strain in the $[111]$-direction, which breaks all the symmetries but time-reversal. The Weyl points survive at small strains, but they move off the high-symmetry $k_z=0$ plane. This should be expected, since the rotational $C_2$ symmetry, which in combination with time-reversal allowed the Weyl points to appear on the high-symmetry plane, is now broken, and a stable Weyl point should appear at arbitrary $k$-points.    

\section{Tight-binding models}
To compute chiralities (Chern numbers) of Weyl points, as well as to interpolate between no SOC and full SOC states, tight-binding models were derived with and without SOC. These tight-binding models were obtained using Wannier interpolation~\cite{Souza-PRB01, Wannier90-2}. Bloch states were projected onto all W $d$-states and all Te $p$-states. We used a specially symmetrized model without SOC that has the same Weyl points distribution as obtained from first-principles, and interpolated it to the one derived from a fully relativistic calculation.

To get the chiralities of Weyl points for the full SOC case, we used an 88-band tight binding model derived from the full SOC first-principles calculation. 

\section{Surface states}

\begin{figure}
\begin{center}
\includegraphics[width=\columnwidth]{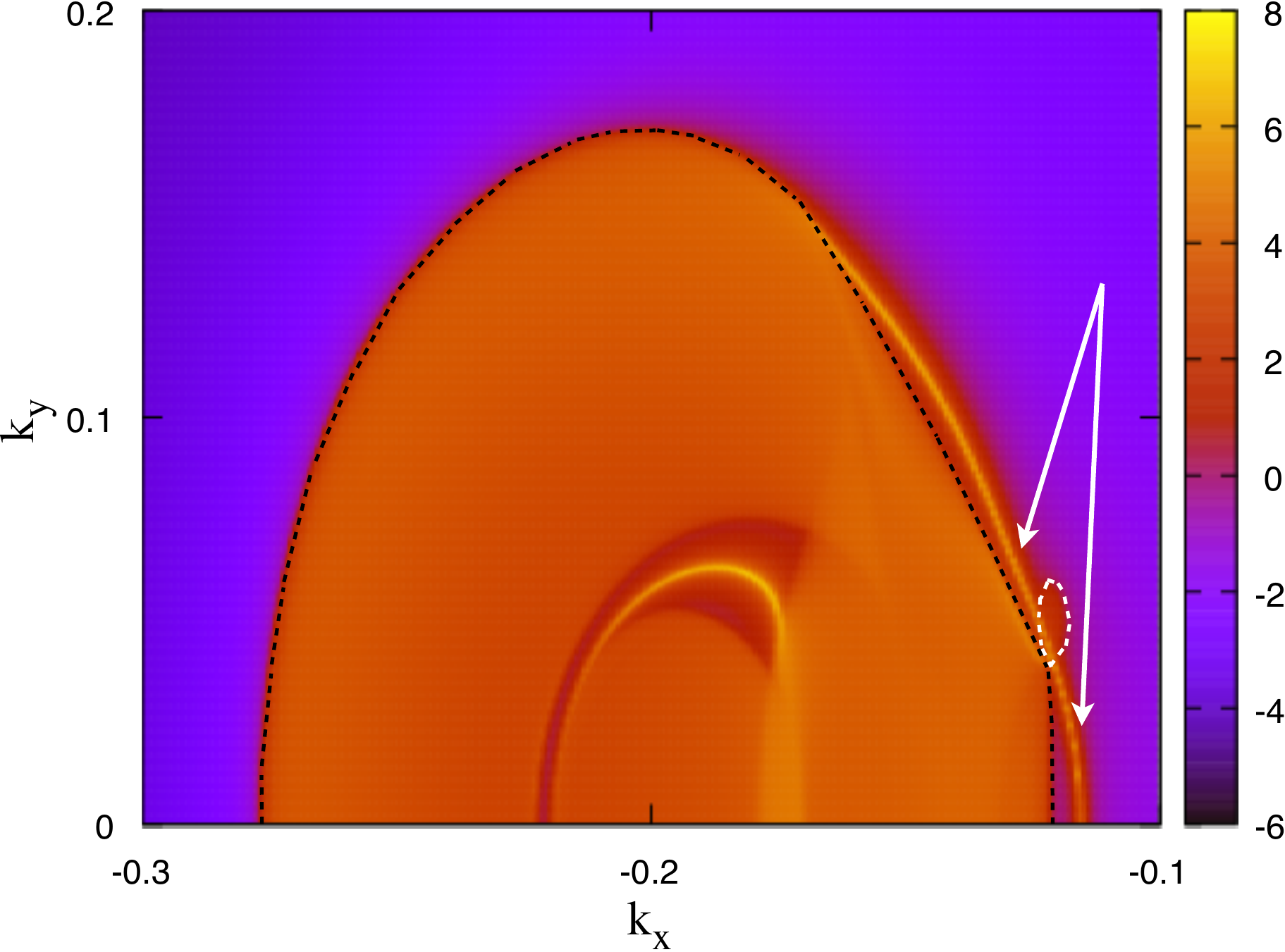}
\end{center}
\caption{Spectral function illustrates the surface Fermi surface. The surface state, indicated with arrows, emerges in the electron pocket (shown with the black contour), crosses the hole pocket (shown with the white contour), and merges back into the electron pocket (by mirror symmetry).}
\label{fig:surfextra}
\end{figure}

Here we provide an illustration to support our assertion, made in the main text, about the possibility to move the Fermi arc that crosses the hole pocket into the continuum of the bulk states. Fig.~\ref{fig:surfextra} shows a larger scale illustration, where it can be seen that this surface state emerges from the electron pocket and merges back into it. Note that unlike the illustration of Fig.~4(b) of the main text, the small electron pocket becomes part of the larger electron pocket in this illustration. This is due to small deviations ($\approx 10$meV) of our tight-binding model band structure from its {\it ab initio} counterpart used for illustrations in Fig.~4 of the main text.

\section{Additional Remarks}
Matplotlib~\cite{Matplotlib}, Mayavi~\cite{mayavi} and VESTA~\cite{vesta}  software packages were used to create some of the illustrations.

\bibliography{paper3} 

\begin{thebibliography}{55}%
\makeatletter
\providecommand \@ifxundefined [1]{%
 \@ifx{#1\undefined}
}%
\providecommand \@ifnum [1]{%
 \ifnum #1\expandafter \@firstoftwo
 \else \expandafter \@secondoftwo
 \fi
}%
\providecommand \@ifx [1]{%
 \ifx #1\expandafter \@firstoftwo
 \else \expandafter \@secondoftwo
 \fi
}%
\providecommand \natexlab [1]{#1}%
\providecommand \enquote  [1]{``#1''}%
\providecommand \bibnamefont  [1]{#1}%
\providecommand \bibfnamefont [1]{#1}%
\providecommand \citenamefont [1]{#1}%
\providecommand \href@noop [0]{\@secondoftwo}%
\providecommand \href [0]{\begingroup \@sanitize@url \@href}%
\providecommand \@href[1]{\@@startlink{#1}\@@href}%
\providecommand \@@href[1]{\endgroup#1\@@endlink}%
\providecommand \@sanitize@url [0]{\catcode `\\12\catcode `\$12\catcode
  `\&12\catcode `\#12\catcode `\^12\catcode `\_12\catcode `\%12\relax}%
\providecommand \@@startlink[1]{}%
\providecommand \@@endlink[0]{}%
\providecommand \url  [0]{\begingroup\@sanitize@url \@url }%
\providecommand \@url [1]{\endgroup\@href {#1}{\urlprefix }}%
\providecommand \urlprefix  [0]{URL }%
\providecommand \Eprint [0]{\href }%
\providecommand \doibase [0]{http://dx.doi.org/}%
\providecommand \selectlanguage [0]{\@gobble}%
\providecommand \bibinfo  [0]{\@secondoftwo}%
\providecommand \bibfield  [0]{\@secondoftwo}%
\providecommand \translation [1]{[#1]}%
\providecommand \BibitemOpen [0]{}%
\providecommand \bibitemStop [0]{}%
\providecommand \bibitemNoStop [0]{.\EOS\space}%
\providecommand \EOS [0]{\spacefactor3000\relax}%
\providecommand \BibitemShut  [1]{\csname bibitem#1\endcsname}%
\let\auto@bib@innerbib\@empty
\bibitem [{\citenamefont {Wan}\ \emph {et~al.}(2011)\citenamefont {Wan},
  \citenamefont {Turner}, \citenamefont {Vishwanath},\ and\ \citenamefont
  {Savrasov}}]{Wan-PRB11}%
  \BibitemOpen
  \bibfield  {author} {\bibinfo {author} {\bibfnamefont {X.}~\bibnamefont
  {Wan}}, \bibinfo {author} {\bibfnamefont {A.~M.}\ \bibnamefont {Turner}},
  \bibinfo {author} {\bibfnamefont {A.}~\bibnamefont {Vishwanath}}, \ and\
  \bibinfo {author} {\bibfnamefont {S.~Y.}\ \bibnamefont {Savrasov}},\
  }\href@noop {} {\bibfield  {journal} {\bibinfo  {journal} {Physical Review
  B}\ }\textbf {\bibinfo {volume} {83}},\ \bibinfo {pages} {205101} (\bibinfo
  {year} {2011})}\BibitemShut {NoStop}%
\bibitem [{\citenamefont {Volovik}(2009)}]{Volovik-book}%
  \BibitemOpen
  \bibfield  {author} {\bibinfo {author} {\bibfnamefont {G.}~\bibnamefont
  {Volovik}},\ }\href@noop {} {\emph {\bibinfo {title} {{The Universe in a
  Helium Droplet}}}}\ (\bibinfo  {publisher} {Oxford University Press New
  York},\ \bibinfo {year} {2009})\BibitemShut {NoStop}%
\bibitem [{\citenamefont {Weng}\ \emph {et~al.}(2015)\citenamefont {Weng},
  \citenamefont {Fang}, \citenamefont {Fang}, \citenamefont {Bernevig},\ and\
  \citenamefont {Dai}}]{Weng-PRX15}%
  \BibitemOpen
  \bibfield  {author} {\bibinfo {author} {\bibfnamefont {H.}~\bibnamefont
  {Weng}}, \bibinfo {author} {\bibfnamefont {C.}~\bibnamefont {Fang}}, \bibinfo
  {author} {\bibfnamefont {Z.}~\bibnamefont {Fang}}, \bibinfo {author}
  {\bibfnamefont {B.~A.}\ \bibnamefont {Bernevig}}, \ and\ \bibinfo {author}
  {\bibfnamefont {X.}~\bibnamefont {Dai}},\ }\href {\doibase
  10.1103/PhysRevX.5.011029} {\bibfield  {journal} {\bibinfo  {journal} {Phys.
  Rev. X}\ }\textbf {\bibinfo {volume} {5}},\ \bibinfo {pages} {011029}
  (\bibinfo {year} {2015})}\BibitemShut {NoStop}%
\bibitem [{\citenamefont {Huang}\ \emph
  {et~al.}(2015{\natexlab{a}})\citenamefont {Huang}, \citenamefont {Xu},
  \citenamefont {Belopolski}, \citenamefont {Lee}, \citenamefont {Chang},
  \citenamefont {Wang}, \citenamefont {Alidoust}, \citenamefont {Bian},
  \citenamefont {Neupane}, \citenamefont {Bansil} \emph
  {et~al.}}]{Huang-NatComm15}%
  \BibitemOpen
  \bibfield  {author} {\bibinfo {author} {\bibfnamefont {S.-M.}\ \bibnamefont
  {Huang}}, \bibinfo {author} {\bibfnamefont {S.-Y.}\ \bibnamefont {Xu}},
  \bibinfo {author} {\bibfnamefont {I.}~\bibnamefont {Belopolski}}, \bibinfo
  {author} {\bibfnamefont {C.-C.}\ \bibnamefont {Lee}}, \bibinfo {author}
  {\bibfnamefont {G.}~\bibnamefont {Chang}}, \bibinfo {author} {\bibfnamefont
  {B.}~\bibnamefont {Wang}}, \bibinfo {author} {\bibfnamefont {N.}~\bibnamefont
  {Alidoust}}, \bibinfo {author} {\bibfnamefont {G.}~\bibnamefont {Bian}},
  \bibinfo {author} {\bibfnamefont {M.}~\bibnamefont {Neupane}}, \bibinfo
  {author} {\bibfnamefont {A.}~\bibnamefont {Bansil}},  \emph {et~al.},\
  }\href@noop {} {\bibfield  {journal} {\bibinfo  {journal} {Nature Comm.}\
  }\textbf {\bibinfo {volume} {6}},\ \bibinfo {pages} {7373} (\bibinfo {year}
  {2015}{\natexlab{a}})}\BibitemShut {NoStop}%
\bibitem [{\citenamefont {Xu}\ \emph {et~al.}(2015)\citenamefont {Xu},
  \citenamefont {Belopolski}, \citenamefont {Alidoust}, \citenamefont
  {Neupane}, \citenamefont {Bian}, \citenamefont {Zhang}, \citenamefont
  {Sankar}, \citenamefont {Chang}, \citenamefont {Yuan}, \citenamefont {Lee},
  \citenamefont {Huang}, \citenamefont {Zheng}, \citenamefont {Ma},
  \citenamefont {Sanchez}, \citenamefont {Wang}, \citenamefont {Bansil},
  \citenamefont {Chou}, \citenamefont {Shibayev}, \citenamefont {Lin},
  \citenamefont {Jia},\ and\ \citenamefont {Hasan}}]{Xu-Science15}%
  \BibitemOpen
  \bibfield  {author} {\bibinfo {author} {\bibfnamefont {S.-Y.}\ \bibnamefont
  {Xu}}, \bibinfo {author} {\bibfnamefont {I.}~\bibnamefont {Belopolski}},
  \bibinfo {author} {\bibfnamefont {N.}~\bibnamefont {Alidoust}}, \bibinfo
  {author} {\bibfnamefont {M.}~\bibnamefont {Neupane}}, \bibinfo {author}
  {\bibfnamefont {G.}~\bibnamefont {Bian}}, \bibinfo {author} {\bibfnamefont
  {C.}~\bibnamefont {Zhang}}, \bibinfo {author} {\bibfnamefont
  {R.}~\bibnamefont {Sankar}}, \bibinfo {author} {\bibfnamefont
  {G.}~\bibnamefont {Chang}}, \bibinfo {author} {\bibfnamefont
  {Z.}~\bibnamefont {Yuan}}, \bibinfo {author} {\bibfnamefont {C.-C.}\
  \bibnamefont {Lee}}, \bibinfo {author} {\bibfnamefont {S.-M.}\ \bibnamefont
  {Huang}}, \bibinfo {author} {\bibfnamefont {H.}~\bibnamefont {Zheng}},
  \bibinfo {author} {\bibfnamefont {J.}~\bibnamefont {Ma}}, \bibinfo {author}
  {\bibfnamefont {D.~S.}\ \bibnamefont {Sanchez}}, \bibinfo {author}
  {\bibfnamefont {B.}~\bibnamefont {Wang}}, \bibinfo {author} {\bibfnamefont
  {A.}~\bibnamefont {Bansil}}, \bibinfo {author} {\bibfnamefont
  {F.}~\bibnamefont {Chou}}, \bibinfo {author} {\bibfnamefont {P.~P.}\
  \bibnamefont {Shibayev}}, \bibinfo {author} {\bibfnamefont {H.}~\bibnamefont
  {Lin}}, \bibinfo {author} {\bibfnamefont {S.}~\bibnamefont {Jia}}, \ and\
  \bibinfo {author} {\bibfnamefont {M.~Z.}\ \bibnamefont {Hasan}},\ }\href
  {\doibase 10.1126/science.aaa9297} {\bibfield  {journal} {\bibinfo  {journal}
  {Science}\ }\textbf {\bibinfo {volume} {349}},\ \bibinfo {pages} {613}
  (\bibinfo {year} {2015})}\BibitemShut {NoStop}%
\bibitem [{\citenamefont {Lv}\ \emph {et~al.}(2015)\citenamefont {Lv},
  \citenamefont {Weng}, \citenamefont {Fu}, \citenamefont {Wang}, \citenamefont
  {Miao}, \citenamefont {Ma}, \citenamefont {Richard}, \citenamefont {Huang},
  \citenamefont {Zhao}, \citenamefont {Chen}, \citenamefont {Fang},
  \citenamefont {Dai}, \citenamefont {Qian},\ and\ \citenamefont
  {Ding}}]{Lv-PRX15}%
  \BibitemOpen
  \bibfield  {author} {\bibinfo {author} {\bibfnamefont {B.~Q.}\ \bibnamefont
  {Lv}}, \bibinfo {author} {\bibfnamefont {H.~M.}\ \bibnamefont {Weng}},
  \bibinfo {author} {\bibfnamefont {B.~B.}\ \bibnamefont {Fu}}, \bibinfo
  {author} {\bibfnamefont {X.~P.}\ \bibnamefont {Wang}}, \bibinfo {author}
  {\bibfnamefont {H.}~\bibnamefont {Miao}}, \bibinfo {author} {\bibfnamefont
  {J.}~\bibnamefont {Ma}}, \bibinfo {author} {\bibfnamefont {P.}~\bibnamefont
  {Richard}}, \bibinfo {author} {\bibfnamefont {X.~C.}\ \bibnamefont {Huang}},
  \bibinfo {author} {\bibfnamefont {L.~X.}\ \bibnamefont {Zhao}}, \bibinfo
  {author} {\bibfnamefont {G.~F.}\ \bibnamefont {Chen}}, \bibinfo {author}
  {\bibfnamefont {Z.}~\bibnamefont {Fang}}, \bibinfo {author} {\bibfnamefont
  {X.}~\bibnamefont {Dai}}, \bibinfo {author} {\bibfnamefont {T.}~\bibnamefont
  {Qian}}, \ and\ \bibinfo {author} {\bibfnamefont {H.}~\bibnamefont {Ding}},\
  }\href {\doibase 10.1103/PhysRevX.5.031013} {\bibfield  {journal} {\bibinfo
  {journal} {Phys. Rev. X}\ }\textbf {\bibinfo {volume} {5}},\ \bibinfo {pages}
  {031013} (\bibinfo {year} {2015})}\BibitemShut {NoStop}%
\bibitem [{\citenamefont {Silaev}\ and\ \citenamefont
  {Volovik}(2012)}]{Silaev-PRB12}%
  \BibitemOpen
  \bibfield  {author} {\bibinfo {author} {\bibfnamefont {M.}~\bibnamefont
  {Silaev}}\ and\ \bibinfo {author} {\bibfnamefont {G.}~\bibnamefont
  {Volovik}},\ }\href@noop {} {\bibfield  {journal} {\bibinfo  {journal}
  {Physical Review B}\ }\textbf {\bibinfo {volume} {86}},\ \bibinfo {pages}
  {214511} (\bibinfo {year} {2012})}\BibitemShut {NoStop}%
\bibitem [{\citenamefont {Nielsen}\ and\ \citenamefont
  {Ninomiya}(1983)}]{Nielsen-PLB83}%
  \BibitemOpen
  \bibfield  {author} {\bibinfo {author} {\bibfnamefont {H.~B.}\ \bibnamefont
  {Nielsen}}\ and\ \bibinfo {author} {\bibfnamefont {M.}~\bibnamefont
  {Ninomiya}},\ }\href@noop {} {\bibfield  {journal} {\bibinfo  {journal}
  {Physics Letters B}\ }\textbf {\bibinfo {volume} {130}},\ \bibinfo {pages}
  {389} (\bibinfo {year} {1983})}\BibitemShut {NoStop}%
\bibitem [{\citenamefont {Zyuzin}\ and\ \citenamefont
  {Burkov}(2012)}]{Zyuzin-PRB12}%
  \BibitemOpen
  \bibfield  {author} {\bibinfo {author} {\bibfnamefont {A.}~\bibnamefont
  {Zyuzin}}\ and\ \bibinfo {author} {\bibfnamefont {A.}~\bibnamefont
  {Burkov}},\ }\href@noop {} {\bibfield  {journal} {\bibinfo  {journal}
  {Physical Review B}\ }\textbf {\bibinfo {volume} {86}},\ \bibinfo {pages}
  {115133} (\bibinfo {year} {2012})}\BibitemShut {NoStop}%
\bibitem [{\citenamefont {Hosur}\ and\ \citenamefont {Qi}(2013)}]{Hosur-crp13}%
  \BibitemOpen
  \bibfield  {author} {\bibinfo {author} {\bibfnamefont {P.}~\bibnamefont
  {Hosur}}\ and\ \bibinfo {author} {\bibfnamefont {X.}~\bibnamefont {Qi}},\
  }\href@noop {} {\bibfield  {journal} {\bibinfo  {journal} {Comptes Rendus
  Physique}\ }\textbf {\bibinfo {volume} {14}},\ \bibinfo {pages} {857}
  (\bibinfo {year} {2013})}\BibitemShut {NoStop}%
\bibitem [{\citenamefont {Volovik}(2014)}]{Volovik-JETPL14}%
  \BibitemOpen
  \bibfield  {author} {\bibinfo {author} {\bibfnamefont {G.}~\bibnamefont
  {Volovik}},\ }\href {\doibase 10.1134/S002136401324020X} {\bibfield
  {journal} {\bibinfo  {journal} {JETP Letters}\ }\textbf {\bibinfo {volume}
  {98}},\ \bibinfo {pages} {753} (\bibinfo {year} {2014})}\BibitemShut
  {NoStop}%
\bibitem [{\citenamefont {Zhang}\ \emph {et~al.}(2015)\citenamefont {Zhang},
  \citenamefont {Xu}, \citenamefont {Belopolski}, \citenamefont {Yuan},
  \citenamefont {Lin}, \citenamefont {Tong}, \citenamefont {Alidoust},
  \citenamefont {Lee}, \citenamefont {Huang}, \citenamefont {Lin} \emph
  {et~al.}}]{Zhang-arx15}%
  \BibitemOpen
  \bibfield  {author} {\bibinfo {author} {\bibfnamefont {C.}~\bibnamefont
  {Zhang}}, \bibinfo {author} {\bibfnamefont {S.-Y.}\ \bibnamefont {Xu}},
  \bibinfo {author} {\bibfnamefont {I.}~\bibnamefont {Belopolski}}, \bibinfo
  {author} {\bibfnamefont {Z.}~\bibnamefont {Yuan}}, \bibinfo {author}
  {\bibfnamefont {Z.}~\bibnamefont {Lin}}, \bibinfo {author} {\bibfnamefont
  {B.}~\bibnamefont {Tong}}, \bibinfo {author} {\bibfnamefont {N.}~\bibnamefont
  {Alidoust}}, \bibinfo {author} {\bibfnamefont {C.-C.}\ \bibnamefont {Lee}},
  \bibinfo {author} {\bibfnamefont {S.-M.}\ \bibnamefont {Huang}}, \bibinfo
  {author} {\bibfnamefont {H.}~\bibnamefont {Lin}},  \emph {et~al.},\
  }\href@noop {} {\bibfield  {journal} {\bibinfo  {journal} {arXiv preprint
  arXiv:1503.02630}\ } (\bibinfo {year} {2015})}\BibitemShut {NoStop}%
\bibitem [{\citenamefont {Xiong}\ \emph {et~al.}(2015)\citenamefont {Xiong},
  \citenamefont {Kushwaha}, \citenamefont {Liang}, \citenamefont {Krizan},
  \citenamefont {Wang}, \citenamefont {Cava},\ and\ \citenamefont
  {Ong}}]{Xiong-arx15}%
  \BibitemOpen
  \bibfield  {author} {\bibinfo {author} {\bibfnamefont {J.}~\bibnamefont
  {Xiong}}, \bibinfo {author} {\bibfnamefont {S.~K.}\ \bibnamefont {Kushwaha}},
  \bibinfo {author} {\bibfnamefont {T.}~\bibnamefont {Liang}}, \bibinfo
  {author} {\bibfnamefont {J.~W.}\ \bibnamefont {Krizan}}, \bibinfo {author}
  {\bibfnamefont {W.}~\bibnamefont {Wang}}, \bibinfo {author} {\bibfnamefont
  {R.}~\bibnamefont {Cava}}, \ and\ \bibinfo {author} {\bibfnamefont
  {N.}~\bibnamefont {Ong}},\ }\href@noop {} {\bibfield  {journal} {\bibinfo
  {journal} {arXiv preprint arXiv:1503.08179}\ } (\bibinfo {year}
  {2015})}\BibitemShut {NoStop}%
\bibitem [{\citenamefont {Huang}\ \emph
  {et~al.}(2015{\natexlab{b}})\citenamefont {Huang}, \citenamefont {Zhao},
  \citenamefont {Long}, \citenamefont {Wang}, \citenamefont {Chen},
  \citenamefont {Yang}, \citenamefont {Liang}, \citenamefont {Xue},
  \citenamefont {Weng}, \citenamefont {Fang}, \citenamefont {Dai},\ and\
  \citenamefont {Chen}}]{Huang-arx15b}%
  \BibitemOpen
  \bibfield  {author} {\bibinfo {author} {\bibfnamefont {X.}~\bibnamefont
  {Huang}}, \bibinfo {author} {\bibfnamefont {L.}~\bibnamefont {Zhao}},
  \bibinfo {author} {\bibfnamefont {Y.}~\bibnamefont {Long}}, \bibinfo {author}
  {\bibfnamefont {P.}~\bibnamefont {Wang}}, \bibinfo {author} {\bibfnamefont
  {D.}~\bibnamefont {Chen}}, \bibinfo {author} {\bibfnamefont {Z.}~\bibnamefont
  {Yang}}, \bibinfo {author} {\bibfnamefont {H.}~\bibnamefont {Liang}},
  \bibinfo {author} {\bibfnamefont {M.}~\bibnamefont {Xue}}, \bibinfo {author}
  {\bibfnamefont {H.}~\bibnamefont {Weng}}, \bibinfo {author} {\bibfnamefont
  {Z.}~\bibnamefont {Fang}}, \bibinfo {author} {\bibfnamefont {X.}~\bibnamefont
  {Dai}}, \ and\ \bibinfo {author} {\bibfnamefont {G.}~\bibnamefont {Chen}},\
  }\href {\doibase 10.1103/PhysRevX.5.031023} {\bibfield  {journal} {\bibinfo
  {journal} {Phys. Rev. X}\ }\textbf {\bibinfo {volume} {5}},\ \bibinfo {pages}
  {031023} (\bibinfo {year} {2015}{\natexlab{b}})}\BibitemShut {NoStop}%
\bibitem [{\citenamefont {Xu}\ \emph {et~al.}(2011)\citenamefont {Xu},
  \citenamefont {Weng}, \citenamefont {Wang}, \citenamefont {Dai},\ and\
  \citenamefont {Fang}}]{Xu-PRL11}%
  \BibitemOpen
  \bibfield  {author} {\bibinfo {author} {\bibfnamefont {G.}~\bibnamefont
  {Xu}}, \bibinfo {author} {\bibfnamefont {H.}~\bibnamefont {Weng}}, \bibinfo
  {author} {\bibfnamefont {Z.}~\bibnamefont {Wang}}, \bibinfo {author}
  {\bibfnamefont {X.}~\bibnamefont {Dai}}, \ and\ \bibinfo {author}
  {\bibfnamefont {Z.}~\bibnamefont {Fang}},\ }\href@noop {} {\bibfield
  {journal} {\bibinfo  {journal} {Physical review letters}\ }\textbf {\bibinfo
  {volume} {107}},\ \bibinfo {pages} {186806} (\bibinfo {year}
  {2011})}\BibitemShut {NoStop}%
\bibitem [{\citenamefont {Burkov}\ and\ \citenamefont
  {Balents}(2011)}]{Burkov-PRL11}%
  \BibitemOpen
  \bibfield  {author} {\bibinfo {author} {\bibfnamefont {A.}~\bibnamefont
  {Burkov}}\ and\ \bibinfo {author} {\bibfnamefont {L.}~\bibnamefont
  {Balents}},\ }\href@noop {} {\bibfield  {journal} {\bibinfo  {journal}
  {Physical review letters}\ }\textbf {\bibinfo {volume} {107}},\ \bibinfo
  {pages} {127205} (\bibinfo {year} {2011})}\BibitemShut {NoStop}%
\bibitem [{\citenamefont {Ali}\ \emph {et~al.}(2014)\citenamefont {Ali},
  \citenamefont {Xiong}, \citenamefont {Flynn}, \citenamefont {Tao},
  \citenamefont {Gibson}, \citenamefont {Schoop}, \citenamefont {Liang},
  \citenamefont {Haldolaarachchige}, \citenamefont {Hirschberger},
  \citenamefont {Ong} \emph {et~al.}}]{Ali-Nature14}%
  \BibitemOpen
  \bibfield  {author} {\bibinfo {author} {\bibfnamefont {M.~N.}\ \bibnamefont
  {Ali}}, \bibinfo {author} {\bibfnamefont {J.}~\bibnamefont {Xiong}}, \bibinfo
  {author} {\bibfnamefont {S.}~\bibnamefont {Flynn}}, \bibinfo {author}
  {\bibfnamefont {J.}~\bibnamefont {Tao}}, \bibinfo {author} {\bibfnamefont
  {Q.~D.}\ \bibnamefont {Gibson}}, \bibinfo {author} {\bibfnamefont {L.~M.}\
  \bibnamefont {Schoop}}, \bibinfo {author} {\bibfnamefont {T.}~\bibnamefont
  {Liang}}, \bibinfo {author} {\bibfnamefont {N.}~\bibnamefont
  {Haldolaarachchige}}, \bibinfo {author} {\bibfnamefont {M.}~\bibnamefont
  {Hirschberger}}, \bibinfo {author} {\bibfnamefont {N.}~\bibnamefont {Ong}},
  \emph {et~al.},\ }\href@noop {} {\bibfield  {journal} {\bibinfo  {journal}
  {Nature}\ }\textbf {\bibinfo {volume} {514}},\ \bibinfo {pages} {205}
  (\bibinfo {year} {2014})}\BibitemShut {NoStop}%
\bibitem [{\citenamefont {Wang}\ \emph {et~al.}()\citenamefont {Wang},
  \citenamefont {Soluyanov}, \citenamefont {Troyer}, \citenamefont {Dai},\ and\
  \citenamefont {Bernevig}}]{genroute}%
  \BibitemOpen
  \bibfield  {author} {\bibinfo {author} {\bibfnamefont {Z.~J.}\ \bibnamefont
  {Wang}}, \bibinfo {author} {\bibfnamefont {A.~A.}\ \bibnamefont {Soluyanov}},
  \bibinfo {author} {\bibfnamefont {M.}~\bibnamefont {Troyer}}, \bibinfo
  {author} {\bibfnamefont {X.}~\bibnamefont {Dai}}, \ and\ \bibinfo {author}
  {\bibfnamefont {B.~A.}\ \bibnamefont {Bernevig}},\ }\href@noop {} {\enquote
  {\bibinfo {title} {{General Route to Topological Semimetals}},}\ }\bibinfo
  {howpublished} {In preparation}\BibitemShut {NoStop}%
\bibitem [{\citenamefont {Soluyanov}\ \emph {et~al.}()\citenamefont
  {Soluyanov}, \citenamefont {Troyer}, \citenamefont {Dai},\ and\ \citenamefont
  {Bernevig}}]{LLs}%
  \BibitemOpen
  \bibfield  {author} {\bibinfo {author} {\bibfnamefont {A.~A.}\ \bibnamefont
  {Soluyanov}}, \bibinfo {author} {\bibfnamefont {M.}~\bibnamefont {Troyer}},
  \bibinfo {author} {\bibfnamefont {X.}~\bibnamefont {Dai}}, \ and\ \bibinfo
  {author} {\bibfnamefont {B.~A.}\ \bibnamefont {Bernevig}},\ }\href@noop {}
  {\enquote {\bibinfo {title} {{Anisotropic Chiral Anomaly in Type-II Weyl
  Semimetals}},}\ }\bibinfo {howpublished} {in preparation}\BibitemShut
  {NoStop}%
\bibitem [{\citenamefont {Nielsen}\ and\ \citenamefont
  {Ninomiya}(1981)}]{Nielsen-NPB81}%
  \BibitemOpen
  \bibfield  {author} {\bibinfo {author} {\bibfnamefont {H.~B.}\ \bibnamefont
  {Nielsen}}\ and\ \bibinfo {author} {\bibfnamefont {M.}~\bibnamefont
  {Ninomiya}},\ }\href@noop {} {\bibfield  {journal} {\bibinfo  {journal}
  {Nuclear Physics B}\ }\textbf {\bibinfo {volume} {185}},\ \bibinfo {pages}
  {20} (\bibinfo {year} {1981})}\BibitemShut {NoStop}%
\bibitem [{\citenamefont {Soluyanov}\ and\ \citenamefont
  {Vanderbilt}(2011{\natexlab{a}})}]{Soluyanov-PRB11-b}%
  \BibitemOpen
  \bibfield  {author} {\bibinfo {author} {\bibfnamefont {A.~A.}\ \bibnamefont
  {Soluyanov}}\ and\ \bibinfo {author} {\bibfnamefont {D.}~\bibnamefont
  {Vanderbilt}},\ }\href@noop {} {\bibfield  {journal} {\bibinfo  {journal}
  {Phys. Rev. B}\ }\textbf {\bibinfo {volume} {83}},\ \bibinfo {pages} {235401}
  (\bibinfo {year} {2011}{\natexlab{a}})}\BibitemShut {NoStop}%
\bibitem [{\citenamefont {Yu}\ \emph {et~al.}(2011)\citenamefont {Yu},
  \citenamefont {Qi}, \citenamefont {Bernevig}, \citenamefont {Fang},\ and\
  \citenamefont {Dai}}]{Yu-PRB11}%
  \BibitemOpen
  \bibfield  {author} {\bibinfo {author} {\bibfnamefont {R.}~\bibnamefont
  {Yu}}, \bibinfo {author} {\bibfnamefont {X.~L.}\ \bibnamefont {Qi}}, \bibinfo
  {author} {\bibfnamefont {A.}~\bibnamefont {Bernevig}}, \bibinfo {author}
  {\bibfnamefont {Z.}~\bibnamefont {Fang}}, \ and\ \bibinfo {author}
  {\bibfnamefont {X.}~\bibnamefont {Dai}},\ }\href@noop {} {\bibfield
  {journal} {\bibinfo  {journal} {Phys. Rev. B}\ }\textbf {\bibinfo {volume}
  {84}},\ \bibinfo {pages} {075119} (\bibinfo {year} {2011})}\BibitemShut
  {NoStop}%
\bibitem [{\citenamefont {Pletikosi\'{c}}\ \emph {et~al.}(2014)\citenamefont
  {Pletikosi\'{c}}, \citenamefont {Ali}, \citenamefont {Fedorov}, \citenamefont
  {Cava},\ and\ \citenamefont {Valla}}]{Pletikosic-PRL14}%
  \BibitemOpen
  \bibfield  {author} {\bibinfo {author} {\bibfnamefont {I.}~\bibnamefont
  {Pletikosi\'{c}}}, \bibinfo {author} {\bibfnamefont {M.~N.}\ \bibnamefont
  {Ali}}, \bibinfo {author} {\bibfnamefont {A.}~\bibnamefont {Fedorov}},
  \bibinfo {author} {\bibfnamefont {R.}~\bibnamefont {Cava}}, \ and\ \bibinfo
  {author} {\bibfnamefont {T.}~\bibnamefont {Valla}},\ }\href {\doibase
  10.1103/PhysRevLett.113.216601} {\bibfield  {journal} {\bibinfo  {journal}
  {Phys. Rev. Lett.}\ }\textbf {\bibinfo {volume} {113}},\ \bibinfo {pages}
  {216601} (\bibinfo {year} {2014})}\BibitemShut {NoStop}%
\bibitem [{\citenamefont {Brown}(1966{\natexlab{a}})}]{Brown-AC66}%
  \BibitemOpen
  \bibfield  {author} {\bibinfo {author} {\bibfnamefont {B.~E.}\ \bibnamefont
  {Brown}},\ }\href@noop {} {\bibfield  {journal} {\bibinfo  {journal} {Acta
  Crystallographica}\ }\textbf {\bibinfo {volume} {20}},\ \bibinfo {pages}
  {268} (\bibinfo {year} {1966}{\natexlab{a}})}\BibitemShut {NoStop}%
\bibitem [{\citenamefont {Goerbig}\ \emph {et~al.}(2008)\citenamefont
  {Goerbig}, \citenamefont {Fuchs}, \citenamefont {Montambaux},\ and\
  \citenamefont {Pi\'echon}}]{Goerbig-PRB08}%
  \BibitemOpen
  \bibfield  {author} {\bibinfo {author} {\bibfnamefont {M.~O.}\ \bibnamefont
  {Goerbig}}, \bibinfo {author} {\bibfnamefont {J.-N.}\ \bibnamefont {Fuchs}},
  \bibinfo {author} {\bibfnamefont {G.}~\bibnamefont {Montambaux}}, \ and\
  \bibinfo {author} {\bibfnamefont {F.}~\bibnamefont {Pi\'echon}},\ }\href
  {\doibase 10.1103/PhysRevB.78.045415} {\bibfield  {journal} {\bibinfo
  {journal} {Phys. Rev. B}\ }\textbf {\bibinfo {volume} {78}},\ \bibinfo
  {pages} {045415} (\bibinfo {year} {2008})}\BibitemShut {NoStop}%
\bibitem [{\citenamefont {Kawarabayashi}\ \emph {et~al.}(2012)\citenamefont
  {Kawarabayashi}, \citenamefont {Hatsugai}, \citenamefont {Morimoto},\ and\
  \citenamefont {Aoki}}]{Kawarabayashi-IJMP12}%
  \BibitemOpen
  \bibfield  {author} {\bibinfo {author} {\bibfnamefont {T.}~\bibnamefont
  {Kawarabayashi}}, \bibinfo {author} {\bibfnamefont {Y.}~\bibnamefont
  {Hatsugai}}, \bibinfo {author} {\bibfnamefont {T.}~\bibnamefont {Morimoto}},
  \ and\ \bibinfo {author} {\bibfnamefont {H.}~\bibnamefont {Aoki}},\ }\href
  {\doibase 10.1142/S2010194512006046} {\bibfield  {journal} {\bibinfo
  {journal} {International Journal of Modern Physics: Conference Series}\
  }\textbf {\bibinfo {volume} {11}},\ \bibinfo {pages} {145} (\bibinfo {year}
  {2012})}\BibitemShut {NoStop}%
\bibitem [{\citenamefont {Xu}\ and\ \citenamefont {Zhang}(2014)}]{Xu-arx14}%
  \BibitemOpen
  \bibfield  {author} {\bibinfo {author} {\bibfnamefont {Y.}~\bibnamefont
  {Xu}}\ and\ \bibinfo {author} {\bibfnamefont {C.}~\bibnamefont {Zhang}},\
  }\href@noop {} {\bibfield  {journal} {\bibinfo  {journal} {arXiv preprint
  arXiv:1411.7316}\ } (\bibinfo {year} {2014})}\BibitemShut {NoStop}%
\bibitem [{\citenamefont {Chang}\ and\ \citenamefont
  {Yang}(2015)}]{Chang-PRB15}%
  \BibitemOpen
  \bibfield  {author} {\bibinfo {author} {\bibfnamefont {M.-C.}\ \bibnamefont
  {Chang}}\ and\ \bibinfo {author} {\bibfnamefont {M.-F.}\ \bibnamefont
  {Yang}},\ }\href {\doibase 10.1103/PhysRevB.91.115203} {\bibfield  {journal}
  {\bibinfo  {journal} {Phys. Rev. B}\ }\textbf {\bibinfo {volume} {91}},\
  \bibinfo {pages} {115203} (\bibinfo {year} {2015})}\BibitemShut {NoStop}%
\bibitem [{\citenamefont {Trescher}\ \emph {et~al.}(2015)\citenamefont
  {Trescher}, \citenamefont {Sbierski}, \citenamefont {Brouwer},\ and\
  \citenamefont {Bergholtz}}]{Trescher-PRB15}%
  \BibitemOpen
  \bibfield  {author} {\bibinfo {author} {\bibfnamefont {M.}~\bibnamefont
  {Trescher}}, \bibinfo {author} {\bibfnamefont {B.}~\bibnamefont {Sbierski}},
  \bibinfo {author} {\bibfnamefont {P.~W.}\ \bibnamefont {Brouwer}}, \ and\
  \bibinfo {author} {\bibfnamefont {E.~J.}\ \bibnamefont {Bergholtz}},\ }\href
  {\doibase 10.1103/PhysRevB.91.115135} {\bibfield  {journal} {\bibinfo
  {journal} {Phys. Rev. B}\ }\textbf {\bibinfo {volume} {91}},\ \bibinfo
  {pages} {115135} (\bibinfo {year} {2015})}\BibitemShut {NoStop}%
\bibitem [{\citenamefont {Hilbert}\ and\ \citenamefont
  {Cohn-Vossen}(1999)}]{Hilbert-book}%
  \BibitemOpen
  \bibfield  {author} {\bibinfo {author} {\bibfnamefont {D.}~\bibnamefont
  {Hilbert}}\ and\ \bibinfo {author} {\bibfnamefont {S.}~\bibnamefont
  {Cohn-Vossen}},\ }\href@noop {} {\emph {\bibinfo {title} {{Geometry and the
  Imagination}}}},\ Vol.~\bibinfo {volume} {87}\ (\bibinfo  {publisher}
  {American Mathematical Soc.},\ \bibinfo {year} {1999})\BibitemShut {NoStop}%
\bibitem [{\citenamefont {Volovik}(1986)}]{Volovik-JETPL86}%
  \BibitemOpen
  \bibfield  {author} {\bibinfo {author} {\bibfnamefont {G.}~\bibnamefont
  {Volovik}},\ }\href@noop {} {\bibfield  {journal} {\bibinfo  {journal} {JETP
  Lett}\ }\textbf {\bibinfo {volume} {43}} (\bibinfo {year}
  {1986})}\BibitemShut {NoStop}%
\bibitem [{\citenamefont {Bevan}\ \emph {et~al.}(1997)\citenamefont {Bevan},
  \citenamefont {Manninen}, \citenamefont {Cook}, \citenamefont {Hook},
  \citenamefont {Hall}, \citenamefont {Vachaspati},\ and\ \citenamefont
  {Volovik}}]{Volovik-Nature97}%
  \BibitemOpen
  \bibfield  {author} {\bibinfo {author} {\bibfnamefont {T.}~\bibnamefont
  {Bevan}}, \bibinfo {author} {\bibfnamefont {A.}~\bibnamefont {Manninen}},
  \bibinfo {author} {\bibfnamefont {J.}~\bibnamefont {Cook}}, \bibinfo {author}
  {\bibfnamefont {J.}~\bibnamefont {Hook}}, \bibinfo {author} {\bibfnamefont
  {H.}~\bibnamefont {Hall}}, \bibinfo {author} {\bibfnamefont {T.}~\bibnamefont
  {Vachaspati}}, \ and\ \bibinfo {author} {\bibfnamefont {G.}~\bibnamefont
  {Volovik}},\ }\href@noop {} {\bibfield  {journal} {\bibinfo  {journal}
  {Nature}\ }\textbf {\bibinfo {volume} {386}},\ \bibinfo {pages} {689}
  (\bibinfo {year} {1997})}\BibitemShut {NoStop}%
\bibitem [{\citenamefont {Volovik}(1998)}]{Volovik-PhysicaB98}%
  \BibitemOpen
  \bibfield  {author} {\bibinfo {author} {\bibfnamefont {G.}~\bibnamefont
  {Volovik}},\ }\href@noop {} {\bibfield  {journal} {\bibinfo  {journal}
  {Physica B: Condensed Matter}\ }\textbf {\bibinfo {volume} {255}},\ \bibinfo
  {pages} {86} (\bibinfo {year} {1998})}\BibitemShut {NoStop}%
\bibitem [{\citenamefont {Son}\ and\ \citenamefont {Spivak}(2013)}]{Son-PRB13}%
  \BibitemOpen
  \bibfield  {author} {\bibinfo {author} {\bibfnamefont {D.}~\bibnamefont
  {Son}}\ and\ \bibinfo {author} {\bibfnamefont {B.}~\bibnamefont {Spivak}},\
  }\href@noop {} {\bibfield  {journal} {\bibinfo  {journal} {Physical Review
  B}\ }\textbf {\bibinfo {volume} {88}},\ \bibinfo {pages} {104412} (\bibinfo
  {year} {2013})}\BibitemShut {NoStop}%
\bibitem [{\citenamefont {Liu}\ \emph {et~al.}(2013)\citenamefont {Liu},
  \citenamefont {Ye},\ and\ \citenamefont {Qi}}]{Liu-PRB13}%
  \BibitemOpen
  \bibfield  {author} {\bibinfo {author} {\bibfnamefont {C.-X.}\ \bibnamefont
  {Liu}}, \bibinfo {author} {\bibfnamefont {P.}~\bibnamefont {Ye}}, \ and\
  \bibinfo {author} {\bibfnamefont {X.-L.}\ \bibnamefont {Qi}},\ }\href@noop {}
  {\bibfield  {journal} {\bibinfo  {journal} {Physical Review B}\ }\textbf
  {\bibinfo {volume} {87}},\ \bibinfo {pages} {235306} (\bibinfo {year}
  {2013})}\BibitemShut {NoStop}%
\bibitem [{\citenamefont {Kharzeev}\ and\ \citenamefont
  {Yee}(2013)}]{Kharzeev-PRB13}%
  \BibitemOpen
  \bibfield  {author} {\bibinfo {author} {\bibfnamefont {D.~E.}\ \bibnamefont
  {Kharzeev}}\ and\ \bibinfo {author} {\bibfnamefont {H.-U.}\ \bibnamefont
  {Yee}},\ }\href@noop {} {\bibfield  {journal} {\bibinfo  {journal} {Physical
  Review B}\ }\textbf {\bibinfo {volume} {88}},\ \bibinfo {pages} {115119}
  (\bibinfo {year} {2013})}\BibitemShut {NoStop}%
\bibitem [{\citenamefont {Burkov}(2014)}]{Burkov-PRL14}%
  \BibitemOpen
  \bibfield  {author} {\bibinfo {author} {\bibfnamefont {A.}~\bibnamefont
  {Burkov}},\ }\href@noop {} {\bibfield  {journal} {\bibinfo  {journal}
  {Physical review letters}\ }\textbf {\bibinfo {volume} {113}},\ \bibinfo
  {pages} {247203} (\bibinfo {year} {2014})}\BibitemShut {NoStop}%
\bibitem [{\citenamefont {Adler}(1969)}]{Adler-PR69}%
  \BibitemOpen
  \bibfield  {author} {\bibinfo {author} {\bibfnamefont {S.~L.}\ \bibnamefont
  {Adler}},\ }\href@noop {} {\bibfield  {journal} {\bibinfo  {journal}
  {Physical Review}\ }\textbf {\bibinfo {volume} {177}},\ \bibinfo {pages}
  {2426} (\bibinfo {year} {1969})}\BibitemShut {NoStop}%
\bibitem [{\citenamefont {Bell}\ and\ \citenamefont
  {Jackiw}(1969)}]{Bell-Cimento69}%
  \BibitemOpen
  \bibfield  {author} {\bibinfo {author} {\bibfnamefont {J.~S.}\ \bibnamefont
  {Bell}}\ and\ \bibinfo {author} {\bibfnamefont {R.}~\bibnamefont {Jackiw}},\
  }\href@noop {} {\bibfield  {journal} {\bibinfo  {journal} {Il Nuovo Cimento
  A}\ }\textbf {\bibinfo {volume} {60}},\ \bibinfo {pages} {47} (\bibinfo
  {year} {1969})}\BibitemShut {NoStop}%
\bibitem [{\citenamefont {Brown}(1966{\natexlab{b}})}]{brownwte2}%
  \BibitemOpen
  \bibfield  {author} {\bibinfo {author} {\bibfnamefont {B.~E.}\ \bibnamefont
  {Brown}},\ }\href@noop {} {\bibfield  {journal} {\bibinfo  {journal} {Acta
  Crystallographica}\ }\textbf {\bibinfo {volume} {20}},\ \bibinfo {pages}
  {268} (\bibinfo {year} {1966}{\natexlab{b}})}\BibitemShut {NoStop}%
\bibitem [{\citenamefont {Mar}\ \emph {et~al.}(1992)\citenamefont {Mar},
  \citenamefont {Jobic},\ and\ \citenamefont {Ibers}}]{wte2latt}%
  \BibitemOpen
  \bibfield  {author} {\bibinfo {author} {\bibfnamefont {A.}~\bibnamefont
  {Mar}}, \bibinfo {author} {\bibfnamefont {S.}~\bibnamefont {Jobic}}, \ and\
  \bibinfo {author} {\bibfnamefont {J.~A.}\ \bibnamefont {Ibers}},\ }\href@noop
  {} {\bibfield  {journal} {\bibinfo  {journal} {Journal of the American
  Chemical Society}\ }\textbf {\bibinfo {volume} {114}},\ \bibinfo {pages}
  {8963} (\bibinfo {year} {1992})}\BibitemShut {NoStop}%
\bibitem [{\citenamefont {Kresse}\ and\ \citenamefont
  {Furthm{\"u}ller}(1996)}]{VASP}%
  \BibitemOpen
  \bibfield  {author} {\bibinfo {author} {\bibfnamefont {G.}~\bibnamefont
  {Kresse}}\ and\ \bibinfo {author} {\bibfnamefont {J.}~\bibnamefont
  {Furthm{\"u}ller}},\ }\href@noop {} {\bibfield  {journal} {\bibinfo
  {journal} {Computational Materials Science}\ }\textbf {\bibinfo {volume}
  {6}},\ \bibinfo {pages} {15} (\bibinfo {year} {1996})}\BibitemShut {NoStop}%
\bibitem [{\citenamefont {Bl{\"o}chl}(1994)}]{PAW1}%
  \BibitemOpen
  \bibfield  {author} {\bibinfo {author} {\bibfnamefont {P.~E.}\ \bibnamefont
  {Bl{\"o}chl}},\ }\href@noop {} {\bibfield  {journal} {\bibinfo  {journal}
  {Physical Review B}\ }\textbf {\bibinfo {volume} {50}},\ \bibinfo {pages}
  {17953} (\bibinfo {year} {1994})}\BibitemShut {NoStop}%
\bibitem [{\citenamefont {Kresse}\ and\ \citenamefont {Joubert}(1999)}]{PAW2}%
  \BibitemOpen
  \bibfield  {author} {\bibinfo {author} {\bibfnamefont {G.}~\bibnamefont
  {Kresse}}\ and\ \bibinfo {author} {\bibfnamefont {D.}~\bibnamefont
  {Joubert}},\ }\href@noop {} {\bibfield  {journal} {\bibinfo  {journal}
  {Physical Review B}\ }\textbf {\bibinfo {volume} {59}},\ \bibinfo {pages}
  {1758} (\bibinfo {year} {1999})}\BibitemShut {NoStop}%
\bibitem [{\citenamefont {Perdew}\ \emph {et~al.}(1996)\citenamefont {Perdew},
  \citenamefont {Burke},\ and\ \citenamefont {Ernzerhof}}]{PBE}%
  \BibitemOpen
  \bibfield  {author} {\bibinfo {author} {\bibfnamefont {J.~P.}\ \bibnamefont
  {Perdew}}, \bibinfo {author} {\bibfnamefont {K.}~\bibnamefont {Burke}}, \
  and\ \bibinfo {author} {\bibfnamefont {M.}~\bibnamefont {Ernzerhof}},\
  }\href@noop {} {\bibfield  {journal} {\bibinfo  {journal} {Physical review
  letters}\ }\textbf {\bibinfo {volume} {77}},\ \bibinfo {pages} {3865}
  (\bibinfo {year} {1996})}\BibitemShut {NoStop}%
\bibitem [{\citenamefont {Kane}\ and\ \citenamefont
  {Mele}(2005)}]{Kane-PRL05-b}%
  \BibitemOpen
  \bibfield  {author} {\bibinfo {author} {\bibfnamefont {C.~L.}\ \bibnamefont
  {Kane}}\ and\ \bibinfo {author} {\bibfnamefont {E.~J.}\ \bibnamefont
  {Mele}},\ }\href@noop {} {\bibfield  {journal} {\bibinfo  {journal} {Phys.
  Rev. Lett.}\ }\textbf {\bibinfo {volume} {95}},\ \bibinfo {pages} {146802}
  (\bibinfo {year} {2005})}\BibitemShut {NoStop}%
\bibitem [{\citenamefont {Fu}\ and\ \citenamefont {Kane}(2006)}]{Fu-PRB06}%
  \BibitemOpen
  \bibfield  {author} {\bibinfo {author} {\bibfnamefont {L.}~\bibnamefont
  {Fu}}\ and\ \bibinfo {author} {\bibfnamefont {C.~L.}\ \bibnamefont {Kane}},\
  }\href@noop {} {\bibfield  {journal} {\bibinfo  {journal} {Phys. Rev. B}\
  }\textbf {\bibinfo {volume} {74}},\ \bibinfo {pages} {195312} (\bibinfo
  {year} {2006})}\BibitemShut {NoStop}%
\bibitem [{\citenamefont {Gresch}\ \emph {et~al.}()\citenamefont {Gresch},
  \citenamefont {Soluyanov}, \citenamefont {Aut\'{e}s}, \citenamefont
  {Vanderbilt}, \citenamefont {Bernevig}, \citenamefont {Yazyev},\ and\
  \citenamefont {Troyer}}]{Z2pack}%
  \BibitemOpen
  \bibfield  {author} {\bibinfo {author} {\bibfnamefont {D.}~\bibnamefont
  {Gresch}}, \bibinfo {author} {\bibfnamefont {A.~A.}\ \bibnamefont
  {Soluyanov}}, \bibinfo {author} {\bibfnamefont {G.}~\bibnamefont
  {Aut\'{e}s}}, \bibinfo {author} {\bibfnamefont {D.}~\bibnamefont
  {Vanderbilt}}, \bibinfo {author} {\bibfnamefont {B.~A.}\ \bibnamefont
  {Bernevig}}, \bibinfo {author} {\bibfnamefont {O.}~\bibnamefont {Yazyev}}, \
  and\ \bibinfo {author} {\bibfnamefont {M.}~\bibnamefont {Troyer}},\ }\href
  {http://z2pack.ethz.ch/} {\enquote {\bibinfo {title} {{Universal Framework
  for Computing Topological Invariants of Band Structures and its Numerical
  Implementation -- Z2Pack}},}\ }\bibinfo {howpublished} {in
  preparation}\BibitemShut {NoStop}%
\bibitem [{\citenamefont {Souza}\ \emph {et~al.}(2001)\citenamefont {Souza},
  \citenamefont {Marzari},\ and\ \citenamefont {Vanderbilt}}]{Souza-PRB01}%
  \BibitemOpen
  \bibfield  {author} {\bibinfo {author} {\bibfnamefont {I.}~\bibnamefont
  {Souza}}, \bibinfo {author} {\bibfnamefont {N.}~\bibnamefont {Marzari}}, \
  and\ \bibinfo {author} {\bibfnamefont {D.}~\bibnamefont {Vanderbilt}},\
  }\href@noop {} {\bibfield  {journal} {\bibinfo  {journal} {Phys. Rev. B}\
  }\textbf {\bibinfo {volume} {65}},\ \bibinfo {pages} {035109} (\bibinfo
  {year} {2001})}\BibitemShut {NoStop}%
\bibitem [{\citenamefont {Mostofi}\ \emph {et~al.}(2014)\citenamefont
  {Mostofi}, \citenamefont {Yates}, \citenamefont {Pizzi}, \citenamefont {Lee},
  \citenamefont {Souza}, \citenamefont {Vanderbilt},\ and\ \citenamefont
  {Marzari}}]{Wannier90-2}%
  \BibitemOpen
  \bibfield  {author} {\bibinfo {author} {\bibfnamefont {A.~A.}\ \bibnamefont
  {Mostofi}}, \bibinfo {author} {\bibfnamefont {J.~R.}\ \bibnamefont {Yates}},
  \bibinfo {author} {\bibfnamefont {G.}~\bibnamefont {Pizzi}}, \bibinfo
  {author} {\bibfnamefont {Y.-S.}\ \bibnamefont {Lee}}, \bibinfo {author}
  {\bibfnamefont {I.}~\bibnamefont {Souza}}, \bibinfo {author} {\bibfnamefont
  {D.}~\bibnamefont {Vanderbilt}}, \ and\ \bibinfo {author} {\bibfnamefont
  {N.}~\bibnamefont {Marzari}},\ }\href {\doibase
  http://dx.doi.org/10.1016/j.cpc.2014.05.003} {\bibfield  {journal} {\bibinfo
  {journal} {Computer Physics Communications}\ }\textbf {\bibinfo {volume}
  {185}},\ \bibinfo {pages} {2309 } (\bibinfo {year} {2014})}\BibitemShut
  {NoStop}%
\bibitem [{\citenamefont {Soluyanov}\ and\ \citenamefont
  {Vanderbilt}(2011{\natexlab{b}})}]{Soluyanov-PRB11-a}%
  \BibitemOpen
  \bibfield  {author} {\bibinfo {author} {\bibfnamefont {A.~A.}\ \bibnamefont
  {Soluyanov}}\ and\ \bibinfo {author} {\bibfnamefont {D.}~\bibnamefont
  {Vanderbilt}},\ }\href@noop {} {\bibfield  {journal} {\bibinfo  {journal}
  {Phys. Rev. B}\ }\textbf {\bibinfo {volume} {83}},\ \bibinfo {pages} {035108}
  (\bibinfo {year} {2011}{\natexlab{b}})}\BibitemShut {NoStop}%
\bibitem [{\citenamefont {King-Smith}\ and\ \citenamefont
  {Vanderbilt}(1993)}]{King-Smith-PRB93}%
  \BibitemOpen
  \bibfield  {author} {\bibinfo {author} {\bibfnamefont {R.~D.}\ \bibnamefont
  {King-Smith}}\ and\ \bibinfo {author} {\bibfnamefont {D.}~\bibnamefont
  {Vanderbilt}},\ }\href@noop {} {\bibfield  {journal} {\bibinfo  {journal}
  {Phys. Rev. B}\ }\textbf {\bibinfo {volume} {47}},\ \bibinfo {pages} {1651}
  (\bibinfo {year} {1993})}\BibitemShut {NoStop}%
\bibitem [{\citenamefont {Hunter}(2007)}]{Matplotlib}%
  \BibitemOpen
  \bibfield  {author} {\bibinfo {author} {\bibfnamefont {J.~D.}\ \bibnamefont
  {Hunter}},\ }\href@noop {} {\bibfield  {journal} {\bibinfo  {journal}
  {Computing in science and engineering}\ }\textbf {\bibinfo {volume} {9}},\
  \bibinfo {pages} {90} (\bibinfo {year} {2007})}\BibitemShut {NoStop}%
\bibitem [{\citenamefont {Ramachandran}\ and\ \citenamefont
  {Varoquaux}(2011)}]{mayavi}%
  \BibitemOpen
  \bibfield  {author} {\bibinfo {author} {\bibfnamefont {P.}~\bibnamefont
  {Ramachandran}}\ and\ \bibinfo {author} {\bibfnamefont {G.}~\bibnamefont
  {Varoquaux}},\ }\href@noop {} {\bibfield  {journal} {\bibinfo  {journal}
  {Computing in Science \& Engineering}\ }\textbf {\bibinfo {volume} {13}},\
  \bibinfo {pages} {40} (\bibinfo {year} {2011})}\BibitemShut {NoStop}%
\bibitem [{\citenamefont {Momma}\ and\ \citenamefont {Izumi}(2011)}]{vesta}%
  \BibitemOpen
  \bibfield  {author} {\bibinfo {author} {\bibfnamefont {K.}~\bibnamefont
  {Momma}}\ and\ \bibinfo {author} {\bibfnamefont {F.}~\bibnamefont {Izumi}},\
  }\href@noop {} {\bibfield  {journal} {\bibinfo  {journal} {Journal of Applied
  Crystallography}\ }\textbf {\bibinfo {volume} {44}},\ \bibinfo {pages} {1272}
  (\bibinfo {year} {2011})}\BibitemShut {NoStop}%
\end{thebibliography}%
 

\end{document}